\documentclass[a4paper,11pt]{article}
\usepackage{jheppub} % for details on the use of the package, please
\usepackage[T1]{fontenc} % if needed
\usepackage{amsmath,amsthm}
\usepackage{amssymb,amsfonts}
\usepackage{graphicx}
\usepackage{dsfont}
\usepackage{relsize}
\usepackage{stmaryrd}
\usepackage{lmodern}
\usepackage{slantsc}
\usepackage{scalefnt}
\usepackage{subfigure}
\usepackage{xspace}
\usepackage{boxedminipage}
\usepackage{ifpdf}
\usepackage{multirow}
\usepackage{xifthen}
\usepackage{tensor}
\usepackage{array}
\usepackage{float}
\usepackage{tikz}
\tikzset{every picture/.style={line width=0.75pt}}
\usepackage{color}
\usepackage{diagbox}
\usetikzlibrary{arrows,matrix,calc,scopes,decorations.markings}
\newcommand{\be}{\begin{equation}}
\newcommand{\ee}{\end{equation}}
\newcommand{\bse}{\begin{subequations}}
\newcommand{\ese}{\end{subequations}}
\newcommand{\ket}[1]{|{#1}\rangle}
\newcommand{\bra}[1]{\langle{#1}|}

\newcommand{\bpm}{\begin{pmatrix}}
\newcommand{\epm}{\end{pmatrix}}
\newcommand{\bmm}{\begin{matrix}}
\newcommand{\emm}{\end{matrix}}

\newcommand{\BLvert}{\Biggl\vert\bmm} %the big \vert on the right
\newcommand{\Brangle}{\emm\Biggr\rangle}

\newcommand{\basis}{
\begin{tikzpicture}[x=0.75pt,y=0.75pt,yscale=-1,xscale=1]
\draw    (270,70) -- (270,90) ;
\draw    (270,90) -- (350,90) ;
\draw    (290,90) -- (290,70) ;
\draw    (350,90) -- (350,70) ;
\draw    (310,90) -- (310,70) ;
\draw  [dash pattern={on 1.5pt off 1.5pt}]  (250,90) -- (270,90) ;
\draw  [dash pattern={on 1.5pt off 1.5pt}]  (350,90) -- (370,90) ;
\draw (269,62) node [anchor=north west][inner sep=0.75pt]  [font=\tiny] [align=left] {$\displaystyle \alpha _{1}$};
\draw (289,62) node [anchor=north west][inner sep=0.75pt]  [font=\tiny] [align=left] {$\displaystyle \alpha _{2}$};
\draw (349,62) node [anchor=north west][inner sep=0.75pt]  [font=\tiny] [align=left] {$\displaystyle \alpha _{n}$};
\draw (319,72) node [anchor=north west][inner sep=0.75pt]  [font=\scriptsize] [align=left] {$\displaystyle \cdots $};
\draw (284,92) node [anchor=north] [inner sep=0.75pt]  [font=\tiny] [align=left] {$\displaystyle \beta _{1}$};
\draw (304,92) node [anchor=north] [inner sep=0.75pt]  [font=\tiny] [align=left] {$\displaystyle \beta _{2}$};
\draw (349,92) node [anchor=north] [inner sep=0.75pt]  [font=\tiny] [align=left] {$\displaystyle \beta _{n-1}$};
\draw (249,82) node [anchor=north west][inner sep=0.75pt]  [font=\tiny] [align=left] {$\displaystyle 1$};
\draw (361,82) node [anchor=north west][inner sep=0.75pt]  [font=\tiny] [align=left] {$\displaystyle 1$};
\end{tikzpicture}
}
\newcommand{\braid}{
\begin{tikzpicture}[x=0.75pt,y=0.75pt,yscale=-1,xscale=1]
\draw    (290,160) -- (350,160) ;
\draw    (329.95,130) .. controls (330.23,149.12) and (309.77,139.03) .. (310,160.03) ;
\draw    (309.95,130.03) .. controls (309.88,136.21) and (311.96,139.33) .. (314.81,141.44) ;
\draw    (323.08,145.96) .. controls (325.8,147.6) and (330.07,151.69) .. (330,160) ;
\draw (319,162) node [anchor=north west][inner sep=0.75pt]  [font=\tiny] [align=left] {$\displaystyle \beta _{i}$};
\draw (329,122) node [anchor=north west][inner sep=0.75pt]  [font=\tiny] [align=left] {$\displaystyle \alpha _{i+1}$};
\draw (301,122) node [anchor=north west][inner sep=0.75pt]  [font=\tiny] [align=left] {$\displaystyle \alpha _{i}$};
\end{tikzpicture}
}
\newcommand{\unbraid}{
\begin{tikzpicture}[x=0.75pt,y=0.75pt,yscale=-1,xscale=1]
\draw    (310,150.43) -- (370,150.43) ;
\draw    (330,130) -- (330,150) ;
\draw    (350,130) -- (350,150) ;
\draw (339,152.43) node [anchor=north west][inner sep=0.75pt]  [font=\tiny] [align=left] {$\displaystyle \beta' _{i}$};
\draw (351,122) node [anchor=north west][inner sep=0.75pt]  [font=\tiny] [align=left] {$\displaystyle \alpha _{i+1}$};
\draw (321,122) node [anchor=north west][inner sep=0.75pt]  [font=\tiny] [align=left] {$\displaystyle \alpha _{i}$};
\end{tikzpicture}
}
\newcommand{\smat}{
\begin{tikzpicture}[x=0.75pt,y=0.75pt,yscale=-1,xscale=1]
\draw  [draw opacity=0] (283.43,144.82) .. controls (279.88,148.04) and (275.17,150) .. (270,150) .. controls (258.95,150) and (250,141.05) .. (250,130) .. controls (250,118.95) and (258.95,110) .. (270,110) .. controls (281.05,110) and (290,118.95) .. (290,130) .. controls (290,133.91) and (288.88,137.56) .. (286.94,140.64) -- (270,130) -- cycle ; \draw   (283.43,144.82) .. controls (279.88,148.04) and (275.17,150) .. (270,150) .. controls (258.95,150) and (250,141.05) .. (250,130) .. controls (250,118.95) and (258.95,110) .. (270,110) .. controls (281.05,110) and (290,118.95) .. (290,130) .. controls (290,133.91) and (288.88,137.56) .. (286.94,140.64) ;  
\draw  [draw opacity=0] (286.95,114.84) .. controls (290.45,111.82) and (295.01,110) .. (300,110) .. controls (311.05,110) and (320,118.95) .. (320,130) .. controls (320,141.05) and (311.05,150) .. (300,150) .. controls (288.95,150) and (280,141.05) .. (280,130) .. controls (280,126.01) and (281.17,122.3) .. (283.18,119.18) -- (300,130) -- cycle ; \draw   (286.95,114.84) .. controls (290.45,111.82) and (295.01,110) .. (300,110) .. controls (311.05,110) and (320,118.95) .. (320,130) .. controls (320,141.05) and (311.05,150) .. (300,150) .. controls (288.95,150) and (280,141.05) .. (280,130) .. controls (280,126.01) and (281.17,122.3) .. (283.18,119.18) ;  
\draw    (250,130) -- (251.84,133.83) ;
\draw    (250,130) -- (248.16,133.83) ;
\draw    (280,130) -- (278.16,133.83) ;
\draw    (280,130) -- (281.84,133.83) ;
\draw (251,102) node [anchor=north west][inner sep=0.75pt]  [font=\tiny] [align=left] {$\displaystyle \alpha $};
\draw (309,102) node [anchor=north west][inner sep=0.75pt]  [font=\tiny] [align=left] {$\displaystyle \beta $};
\end{tikzpicture}
}
\newcommand{\loopamp}{\begin{tikzpicture}[x=0.75pt,y=0.75pt,yscale=-1,xscale=1]
\draw    (270,130) .. controls (243.24,129.53) and (242.1,90.1) .. (270,90) ;
\draw    (340,90) .. controls (367.81,90.1) and (367.24,130.1) .. (340,130) ;
\draw    (320,90) -- (340,90) ;
\draw    (320,130) -- (340,130) ;
\draw    (279.71,90) -- (279.71,130) ;
\draw    (330,90) -- (330,130) ;
\draw    (247.24,111.53) -- (249.43,106) ;
\draw    (252.1,111.53) -- (249.43,106) ;
\draw    (279.71,105.71) -- (281.81,110.4) ;
\draw    (330,106.86) -- (332.1,111.54) ;
\draw    (360.86,109.71) -- (362.95,114.4) ;
\draw    (279.71,105.71) -- (277.53,110.4) ;
\draw    (330,106.86) -- (327.53,111.26) ;
\draw    (360.86,109.71) -- (357.81,114.11) ;
\draw    (270,90) -- (290,90) ;
\draw    (270,130) -- (290,130) ;
\draw  [dash pattern={on 1.5pt off 1.5pt}]  (290,90) -- (320,90) ;
\draw  [dash pattern={on 1.5pt off 1.5pt}]  (290,130) -- (320,130) ;
\draw (240.02,117.95) node [anchor=north west][inner sep=0.75pt]  [font=\tiny] [align=left] {$\displaystyle x_{1}$};
\draw (268.3,117.66) node [anchor=north west][inner sep=0.75pt]  [font=\tiny] [align=left] {$\displaystyle x_{2}$};
\draw (341.73,117.38) node [anchor=north west][inner sep=0.75pt]  [font=\tiny] [align=left] {$\displaystyle x_{n}$};
\draw (294.94,102.85) node [anchor=north west][inner sep=0.75pt]  [font=\scriptsize] [align=left] {$\displaystyle \cdots $};
\end{tikzpicture}
}

\title{Characterizing the ambiguity in topological entanglement entropy}
\date{\today}
\author[a]{Yingcheng Li}
\affiliation[a]{State Key Laboratory of Surface Physics, Department of Physics, Center for Field Theory and Particle Physics,
and Institute for Nanoelectronic Devices and Quantum Computing, Fudan University, Shanghai 200433, China}
\emailAdd{liyc5170@gmail.com}

\abstract{
Topological entanglement entropy (TEE), the sub-leading term in the entanglement entropy of topological order, is the direct evidence of the \textit{long-range entanglement}. While effective in characterizing topological orders on closed manifolds, TEE is model-dependent when entanglement cuts intersect with physical gapped boundaries. In this paper, we study the origin of this model-dependence by introducing a model-independent picture of partitioning the topological orders with gapped boundaries. In our picture, the entanglement boundaries (EBs), i.e. the virtual boundaries of each subsystem induced by the entanglement cuts, are assumed to be gapped boundaries with boundary defects. At this model-independent stage, there are two choices one has to make manually in defining the bi-partition: the boundary condition on the EBs, and the coherence between certain boundary states. We show that TEE appears because of a constraint on the defect configurations on the EBs, which is choice-dependent in the cases where the EBs touch gapped boundaries. This choice-dependence is known as the \textit{ambiguity} in entanglement entropy. Different models intrinsically employ different choices, rendering TEE model-dependent. For $D(\mathbb{Z}_2)$ topological order, the ambiguity can be fully characterized by two parameters that respectively quantifies the EB condition and the coherence. In particular, calculations compatible with the \textit{folding trick} naturally choose EB conditions that respect \textit{electric-magnetic duality} and set specific parameter values.
}

\begin{document}
\maketitle
\flushbottom
\section{Introduction}\label{sec:intro}

%chiral for CS, non-chiral for lattice, relation between gapped and gapless boundary? 
%5. Model independence = RG flow invariant

Topologically ordered phases of matter, or topological orders for short, are exotic gapped phases in two spatial dimension that exhibit long-range entanglement\cite{chen_local_2010}. This property of entanglement can be captured by the entanglement entropy of the system. For gapped matter phases in two dimension, their ground states usually have entanglement entropy that yields the area-law, i.e the entanglement entropy is proportional to the length of the entanglement cut; but for the ground state of a topological order on a sphere, there is an extra sub-leading term that survive at arbitrary long range. This sub-leading term is called the \textit{Topological Entanglement Entropy}\cite{kitaev_topological_2006} (TEE).

\begin{figure}[h!]
    \centering
    \includegraphics[scale=0.35]{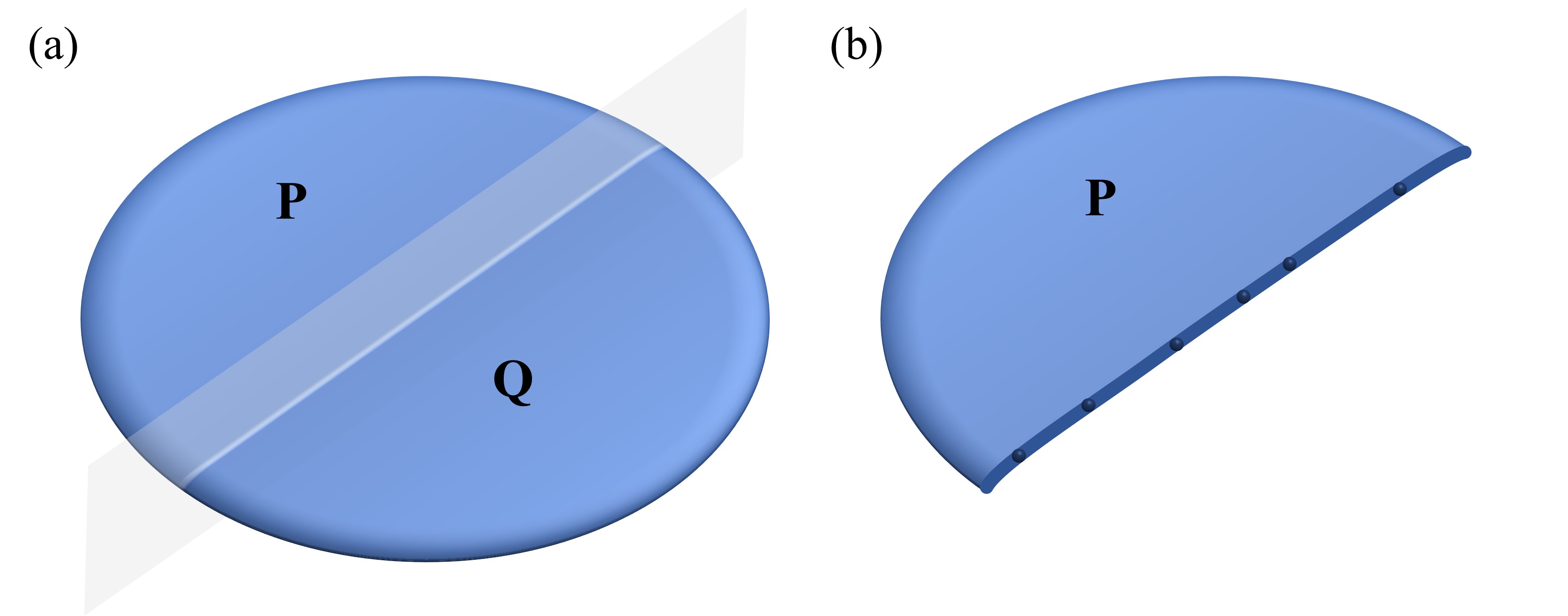}
    \caption{Topological order on a disk with a gapped boundary. (a) The disk is divided into two subsystems, and the entanglement cut intersects with the physical boundary. One calculate the entanglement by partially tracing out either of the two subsystems and calculate the von Neumann entropy of the reduced density matrix. (b) Each subsystem is a disk with two boundary segments: a segment of physical boundary and a segment of the virtual entanglement boundary (thick blue line). In our picture, the entanglement boundaries are gapped boundaries with boundary defects (dark blue dots). }
    \label{fig1:intro}
\end{figure}

Topological entanglement entropy can be systematically and analytically studied using the effective theories of topological orders, which are generally classified into two different kinds: the Chern-Simons (CS) theory \cite{witten_quantum_1989}, including the $K$-matrix theory\cite{wen_quantum_2010,hung_matrix_2013}, and the lattice models such as the Levin-Wen (LW) model\cite{levin_stringnet_2005,hung_stringnet_2012,hu_boundary_2017,hu_boundary_2018,wang_extend_2022} and the quantum double (QD) model\cite{kitaev_faulttolerant_2003,cong_hamiltonian_2017,hu_twisted_2013,bullivant_twisted_2017,hu_full_2018}. In the lattice models, the overall Hilbert space is inherently structured as a tensor product of the Hilbert spaces of local degrees of freedom (d.o.f.), and hence it is straightforward to bi-partition the lattice models and calculate entanglement entropy\cite{levin_detecting_2006,hung_revisiting_2015,luo_quantum_2016,chen_entanglement_2018,hu_entanglement_2019}. In the CS theory, however, one requires certain methods such as the edge theory approach\cite{wen_edge_2016,lou_ishibashi_2019,shen_ishibashi_2019} to derive entanglement entropy. For topological orders on closed manifolds, calculating TEE using different models gives same results. Nevertheless, for topological orders on open manifolds with gapped boundaries, different theories may produce different TEE when the entanglement cuts touch gapped boundaries. For example, consider the $\mathbb{Z}_2$ toric code on a disk with a gapped boundary and bi-partition the disk in the way shown in Fig. \ref{fig1:intro}. The calculation in the QD model shows that TEE varies with the boundary condition of the physical boundary\cite{chen_entanglement_2018}, while the calculation in the $K$-matrix theory shows that TEE is independent of the boundary condition\cite{shen_ishibashi_2019}. It requires clarification what information is encoded in TEE and what leads to the model-dependence in these open-cut cases. 

In this paper, we focus on the topological orders that allow gapped boundaries and study the origin of this model-dependence in Abelian topological orders via a model-independent picture of bi-partitioning the topological order. Suppose we have a topological order on a two-dimensional manifold, and divide the manifold into two subsystems P and Q. For each subsystem, the entanglement cuts serve as virtual entanglement boundaries (EBs), and the Hilbert space of each subsystem is spanned by the boundary states describing EBs. In our picture, the EBs are assumed to be gapped boundaries, and the EB states describe different configurations of boundary defects on the EBs. These boundary defects are anyonic objects that can be model-independently characterized by the bulk topological order and the gapped boundary condition (GBC)\cite{kong_anyon_2014,cong_defects_2017,shen_defect_2019}, rendering the EB states and the corresponding subsystem Hilbert spaces model-independent. The TEE, as we will see in our picture, results from a constraint on each EB that only captures the universal properties of the boundary states. 

Given the subsystem Hilbert spaces, the global state can be decomposed into a summation of tensor product states of the two subsystem Hilbert spaces. At the model-independent level, there are two choices one has to make manually in defining the decomposition: the boundary conditions on the EBs, and the coherence between the boundary states. The former specifies the explicit form of the EB states, while the latter determines the eigenstates and the eigenvalues of the reduced density matrix. This dependence on the manual choices is the \textit{ambiguity} in entanglement entropy\cite{casini_remarks_2014}. The question is, whether different choices change the universal properties of the boundary states? We answer this question in Abelian topological orders by introducing an operator version of the EB constraint and using this operator constraint to pick out the universal properties. It turns out that when the EBs touch gapped boundaries, the TEE will explicitly depend on the EB conditions and the coherence between the two junctions connecting the EB and physical boundaries. Different effective theories have different intrinsic choices of the EB condition and the coherence, rendering TEE model-dependent. Particularly, the lattice models choose the EB conditions that condense specific anyons, while the $K$-matrix theory, which employs the \textit{folding trick}, naturally choose EB conditions that respect \textit{electric-magnetic} (EM) \textit{duality}\cite{wang_electricmagnetic_2020} and set a specific coherence pattern between certain boundary states. For the $\mathbb{Z}_2$ toric code, we introduce a general definition of the decomposition with two parameters that respectively characterize the EB condition and the coherence. These two parameters fully quantifies the ambiguity of TEE in the $\mathbb{Z}_2$ toric code.

Once the ambiguity is characterized, the universal properties encoded in TEE is then transparent. For each EB, there are two individual facts that contribute to TEE: the total topological charge that goes through the EB and the defect species that are allowed on the EB. While the latter exhibits choice-dependnce, the former is choice-independent and contributes a ln$d_x$ term to TEE, where $d_x$ is the quantum dimension of a bulk anyon or a boundary defect in different cases. For a generic state, there will also be a classical term encoding the probability of finding certain basis states, as already recognized in ref \cite{wen_edge_2016}.

This paper is organized as follows. In section 2, we review some useful facts about gapped boundaries and boundary defects of topological orders. We will see the constraint on the boundary defect configurations in a general picture. In section 3, we focus on the closed EB cases and introduce the model-independent picture. We will first demonstrate the procedure of calculating TEE, and then discuss the operator version of the constraint. In section 4, we use our picture to investigate the open EB cases with the simplest choice of EB conditions. In section 5, we will show the general definition of the state decomposition for $\mathbb{Z}_2$ toric code. We will reproduce the result calculated in the CS theory by employing choices that respect EM duality. We comment on some important points and conclude in section 6.

\section{Review of gapped boundaries and boundary defects}\label{sec:review}

In this section, we review some facts about the gapped boundaries of topological orders and clarify the notations that we will use later. A review about the basic properties of anyons can be found in the Appendix \ref{appd:anyon}. For a topological order, we use Greek alphabets $\alpha,\beta,\gamma,\cdots$ to label the species of bulk anyons. The defects on a specific gapped boundary, also called boundary anyons, are labeled by lower Latin alphabets $a,b,c,\cdots$. The trivial anyon is denoted as $1$. The set of bulk anyons is denoted as $\mathcal{C}$, and the set of boundary anyons is denoted as $\mathcal{D}$.

A gapped boundary is characterized by a \textit{Lagrangian subsets} $\mathcal{L}$ of $\mathcal{C}$, i.e. a maximal set of anyons that are self-bosons and have mutually trivial braiding statistics. All the anyons in $\mathcal{L}$ can freely annihilate or create at the boundary. This mechanism is called the \textit{anyon condensation}\cite{kong_anyon_2014}. Other anyons in $\mathcal{C}$ are \textit{confined} at the boundary, and will decompose into boundary anyons if moved onto the boundary. The relation between the bulk anyons $\alpha\in\mathcal{C}$ and the boundary anyons $a\in\mathcal{D}$ can be expressed by a matrix $W$,
\be
\alpha=\bigoplus_{a\in\mathcal{D}} W_{\alpha,a} a, 
\ee
where the matrix elements $W_{\alpha,a}$ are non-negative integers. If $\alpha\in\mathcal{L}$, then $W_{\alpha,1}\neq 0$. This decomposition process commute with fusion, i.e.,
\be
\sum_{\gamma} N_{\alpha\beta}^{\gamma}=\sum_{a,b} N_{ab}^c W_{\alpha,a}W_{\beta,b}.
\ee
For a certain bulk anyon $\beta$, there can be more than one $b$ satisfying $W_{\beta,b}\neq 0$. We say that the anyon $\beta$ \textit{splits} at the boundary. We also have the following two relations for quantum dimensions of the bulk anyons and the boundary anyons,
\begin{align}
D=\sqrt{\sum_{\alpha} d_\alpha^2}=\sum_{a}d_a^2\label{Eq:corresp}\\
d_{\alpha}=\sum_a W_{\alpha,a}d_a\label{Eq:qdim}.
\end{align}

\begin{figure}[h!]
    \centering
    \includegraphics[scale=0.4]{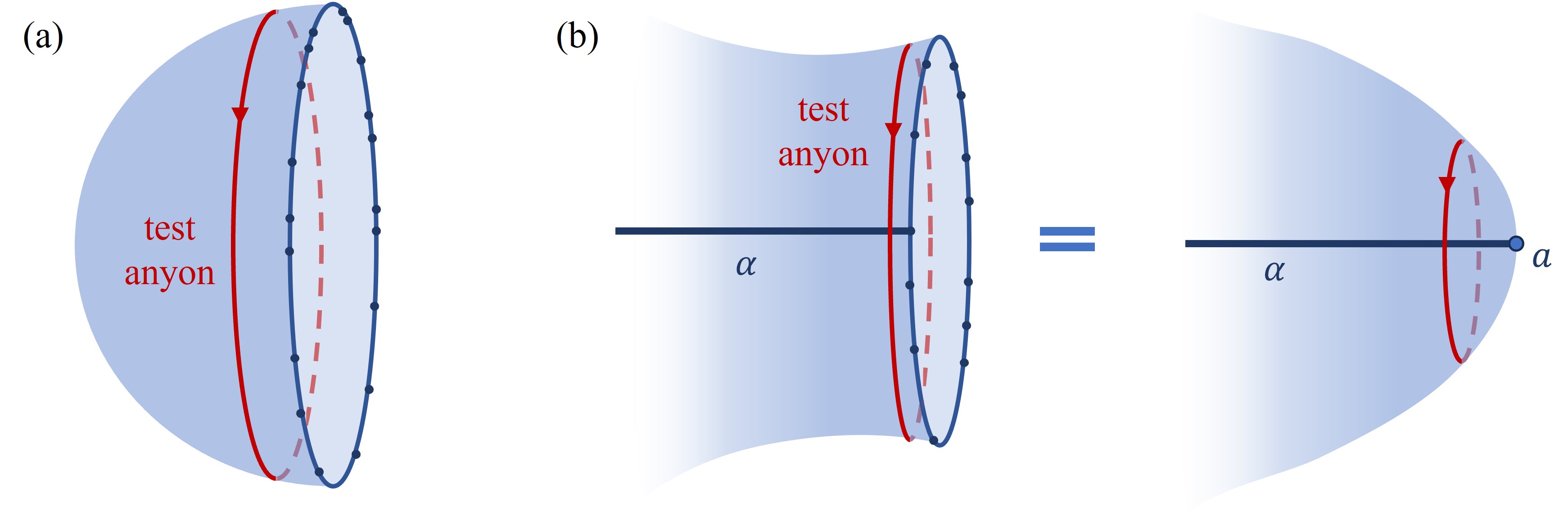}
    \caption{A topological order with a gapped boundary. The blue dots on the boundary represent for boundary anyons. (a) A test anyon loop near the boundary. The anyon loop will find that the boundary has total topological charge $1$. (b) The test anyon loop detect a non-trivial bulk anyon $\alpha$. One can shrink the gapped boundary into a point, during which the boundary anyons are fused to a single anyon $a$. This single anyon $a$ reside at the point should satisfy $W_{\alpha,a}\neq 0$.}
    \label{fig2:closebdry}
\end{figure}

There is a constraint on the configurations of boundary defects on a boundary. To see it, we consider a simple case of a disk with no anyons in the bulk but some anyons on the boundary. We can measure the total topological charge of the boundary by creating a pair of `test anyons', let one of them travel through a trajectory that is very close to the boundary and then annihilate with its partner, as shown in the Fig. \ref{fig2:closebdry}a. This closed anyon loop, as one find immediately, is contractible and detects nothing. Hence, the gapped boundary should have total charge $1$, and consequently all the boundary anyons should fuse to $1$. A general case is depicted in Fig. \ref{fig2:closebdry}b. Suppose the boundary anyons fuse to a non-trivial anyon $a$, then the test anyon loop should recognize the boundary as a bulk anyon $\alpha$ that satisfies $W_{\alpha,a}\neq 0$. In this sense, the gapped boundaries are treated on the same footing as bulk anyons\cite{li_anyonic_2019}. Conversely, if the bulk anyons on a disk fuse to $\alpha$, the boundary total charge should be $a$ that satisfies $W_{\alpha,a}\neq 0$. If $\alpha$ splits on the boundary, the boundary anyons can fuse to any of the $a$'s decomposed from $\alpha$. 

A topological order can have multiple Lagrangian subsets. Each Lagrangian subset corresponds to a certain gapped boundary condition (GBCs). We use upper Latin alphabets $A,B,C,\cdots$ to denote the GBCs of a topological order. The junctions between the boundaries with different GBCs also host anyonic defects. We will call them junction anyons hereinafter. We denote the set of allowed junction anyons between two boundaries with GBC $A$ and $B$ as $\mathcal{D}^{A|B}=\{a^{A|B}, b^{A|B}, ...\}$. The number of elements in $\mathcal{D}^{A|B}$ equals $\sum_\alpha W^A_{\alpha,1^A}W^B_{\alpha,1^B}$, where $W^A$ ($W^B$) is the $W$ matrix on the boundary $A$ ($B$) and $1^A$ ($1^B$) is the trivial anyon on the boundary $A$ ($B$). The junction anyons also subject to fusion rules,
\be
a^{A|B} \times b^{B|C}=\bigoplus_{c^{A|C}} N_{a^{A|B}b^{B|C}}^{c^{A|C}} \hspace{5pt}c^{A|C}
\ee
where $N_{a^{A|B}b^{B|C}}^{c^{A|C}}$'s are non-negative integers. If $A=B$, i.e. the junction connects two boundaries with same GBC, the set of allowed junction anyons then equals the set of boundary anyons $\mathcal{D}^{A|A}=\mathcal{D}^A$. We use the term `boundary defects' to refer to both boundary anyons and junction anyons.

\begin{figure}[h!]
    \centering
    \includegraphics[scale=0.4]{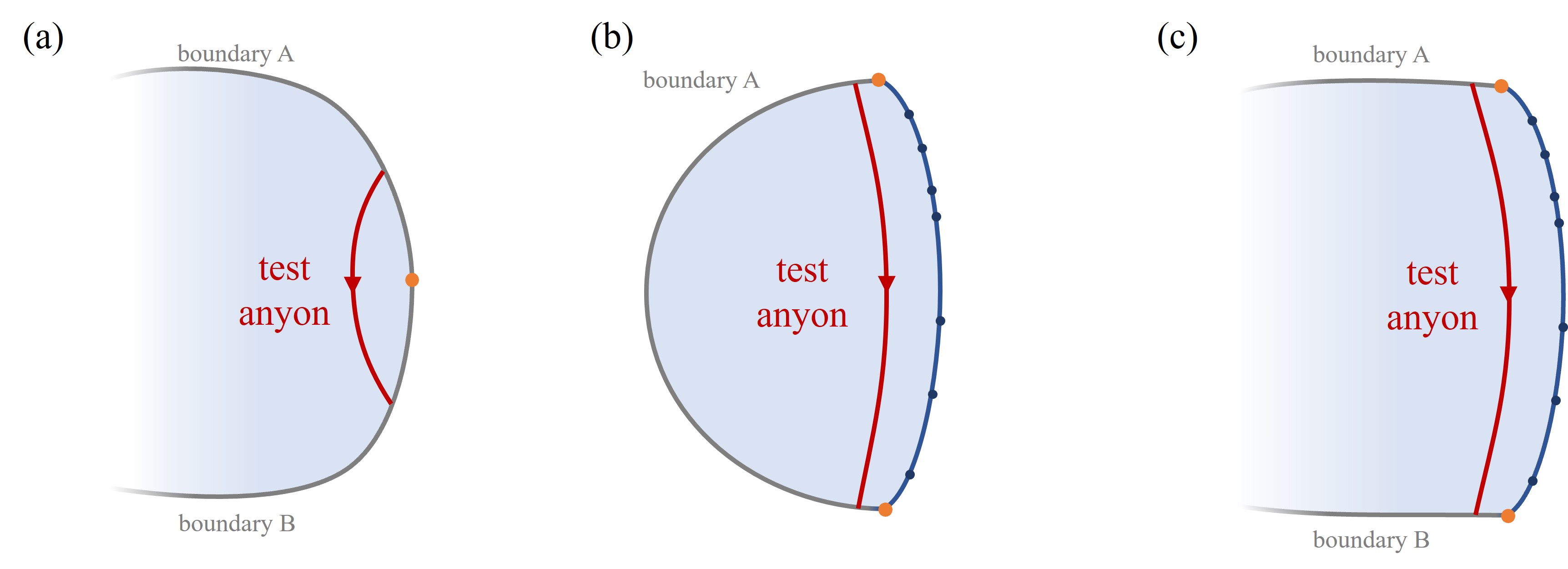}
    \caption{A topological order with multiple boundary segments. The orange dots represent for junctions between boundaries. (a) The half loop that detects the boundary defect. The orange dot represents for a generic
    boundary defect. (b) A test half loop near the boundary. The half loop will find that the boundary has total topological charge $1$. (c) The half loop may detect a non-trivial boundary defect $a^{A|B}$. If we shrink the blue boundary segment into a point, through which the boundary defects fuse to a single defect, we get the same picture as (a).}
    \label{fig3:openbdry}
\end{figure}

Like a bulk anyon can be detected by a bulk anyon loop, a boundary defect can be detected by a half loop. Suppose an anyon $\gamma$ condenses at both $A$ and $B$, and a generic defect $c^{A|B}$ reside on the junction. One can create an $\gamma$ at boundary $A$, let it go around the junction through the bulk, and then annihilate at boundary $B$, as shown in Fig. \ref{fig3:openbdry}a. The trajectory of $\gamma$ forms a half loop, which generates a unitary transformation $U_{\gamma,c^{A|B}}$ on the defect states. This unitary transformation $U_{\gamma,c^{A|B}}$ is like the `braiding' of a condensed anyon and a boundary defect, and is analogous to the braiding of bulk anyons. If we further annihilate the defect with its conjugate counterpart, we will result in the \textit{half-linking} introduced in ref \cite{shen_defect_2019} and \cite{kapustin_topological_2011}. 

There is also a constraint on the configurations of boundary defects on an open boundary segment. Suppose we have a disk with two boundary segments with GBCs $A$ and $B$, and there is no anyon in the bulk or on the boundary $B$. Only boundary $A$ and the two junctions bear anyons. We can then detect the total topological charge of the boundary segment $A$ via the half-loop detection: we create an anyon $\gamma\in\mathcal{L}^B$ on the boundary $B$ near one of the junctions, let $\gamma$ travel very close to boundary $A$ though the bulk and annihilate near the other junction on boundary $B$, as shown in Fig. \ref{fig3:openbdry}b. Again, one find immediately that this half loop is contractible and detects nothing. Therefore, all the defects, including the anyons on boundary $A$ and the two junction anyons, should fuse to
\be
j_0^{B|A}\times x_1^A \times x_2^A\cdots\times j_1^{A|B}=1^B,
\ee
where $j_i^{B|A}$ denotes the anyons reside on the two junctions and $x_i^A$ denotes the anyons on boundary $A$. 

More general cases involve multiple boundary segments, where the boundary anyons and the junction anyons may fuse to a generic boundary defect, as illustrated in Fig. \ref{fig3:openbdry}c. We will see that the constraint on the configurations of boundary defects plays a crucial role in deriving the entanglement entropy and eventually leads to TEE.

Before we proceed, we would like to collect some useful facts about the $\mathbb{Z}_2$ toric code, which we would heavily use in the rest of this paper. There are $4$ kinds of anyons in the $\mathbb{Z}_2$ toric code, which are labelled as $\{1,e,m,\epsilon\}$. The anyon $\epsilon$ can be viewed as a composition of anyon $e$ and $m$. All the four anyons are self-conjugate and have quantum dimension $1$. The remaining non-trivial fusion rules are given by
\be
e\times m= \epsilon;\hspace{15pt} m\times \epsilon=e;\hspace{15pt} e\times\epsilon=m.
\ee
The toric code has two Lagrangian subsets $A=\{1,m\}$ and $B=\{1,e\}$. The bulk anyons $\{e,\epsilon\}$ ($\{m,\epsilon\}$) are identified on the boundary with GBC $A$ ($B$) and become the same boundary anyon $e^A$ ($m^B$). The junction between two boundaries with different GBC can only bear one kind of anyon $\sigma^{A|B}$. The fusion rules yields
\begin{align*}
&\sigma^{A|B}\times m^B= \sigma^{A|B}; \hspace{10pt} \sigma^{B|A}\times e^A= \sigma^{B|A}; \\
&\sigma^{A|B} \times \sigma^{B|A}= 1^A +e^A; \hspace{10pt}
\sigma^{B|A} \times \sigma^{A|B}= 1^B +m^B.
\end{align*}
We have $d_{e^A}=d_{m^B}=1$ and $d_{\sigma^{A|B}}=\sqrt{2}$.

\section{The boundary defect picture: closed EBs}\label{sec:pic}

In this section, we introduce the boundary defect picture in the cases where the EBs are closed loops. In our picture, the EBs are assumed to be gapped boundaries, and the Hilbert space of each subsystem is spanned by the states of different configurations of boundary anyons. We will focus on two cases: the unique ground state of a topological order on a sphere, and the ground states on a torus. The result we will derive in this section is consistent with that in ref \cite{wen_edge_2016}, where the authors used the CS theory to calculate the TEE and discussed general cases. Then, we will discuss the operators of the subsystem for Abelian topological orders. Some contractible anyon loops in the perspective of a global observer are cut by the EBs and become half loops in each subsystem, as shown in Fig. \ref{fig4:bipartsphere}. These half loops will create or annihilate boundary defects on the EBs. There is a constraint on these bi-partitioned operators, which will help us to distinguish the TEE from the local contributions of sub-leading terms.

\subsection{Case: ground state on a sphere}\label{subsec:sphere}
\begin{figure}[h!]
    \centering
    \includegraphics[scale=0.4]{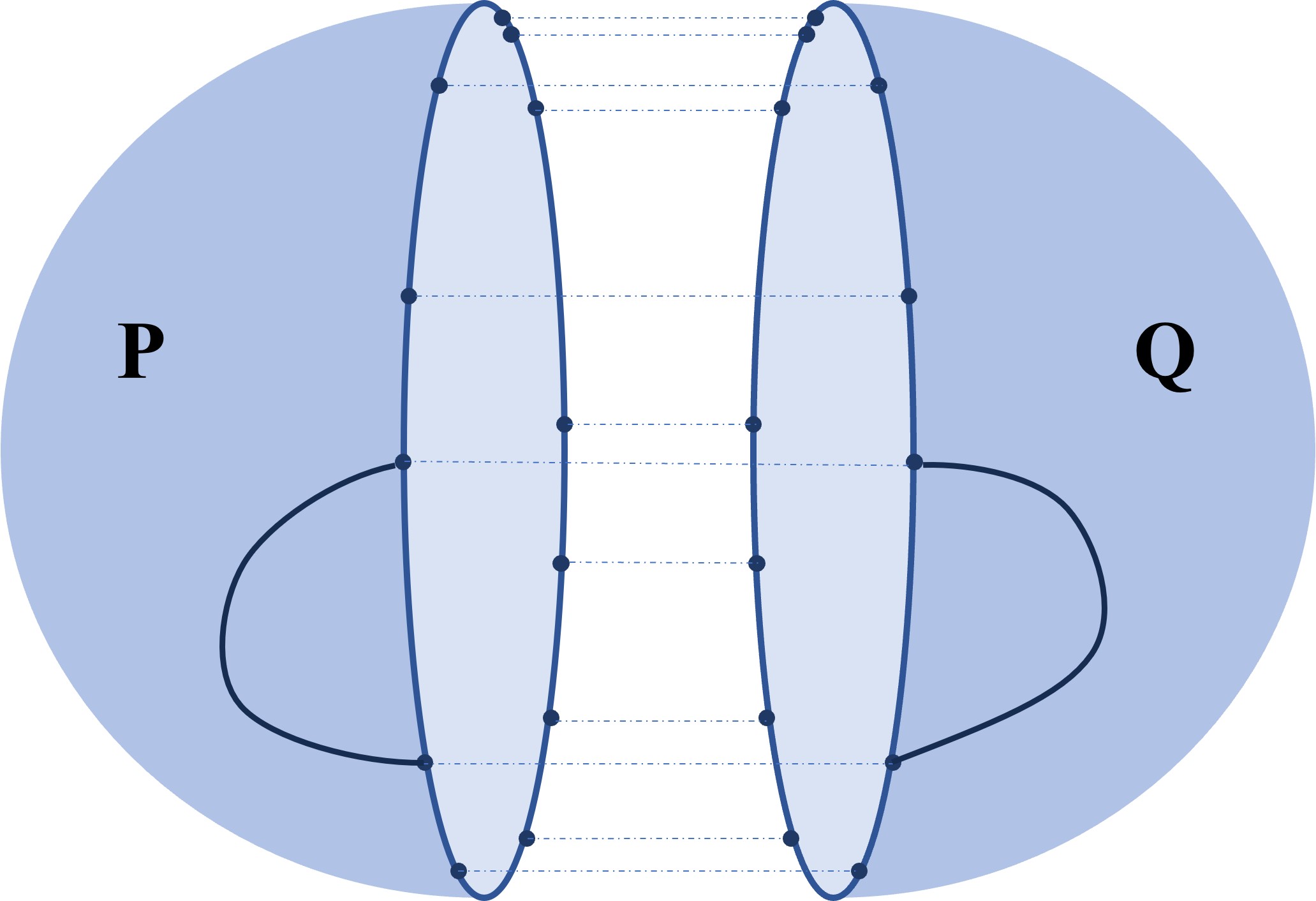}
    \caption{The topological order on a sphere is divided into two subsystems P and Q. We suppose there are $n=L/\delta$ discrete points homogeneously distributed on the EB that can host boundary anyons. A contractible anyon loop (dark blue line) is also divided into two half loops; in the perspective of a subsystem observer, such half loops will create or annihilate boundary anyons on the EB.}
    \label{fig4:bipartsphere}
\end{figure}

We start from the simplest case: the ground state of a topological order on a sphere, as shown in Fig. \ref{fig4:bipartsphere}. We assume that the EB have a fixed GBC and omit the superscript that labels the GBC for the moment. Suppose the length of the EB is $L$, and for simplicity we suppose the points that can host boundary anyons on the EB are discrete and homogeneously distributed on the EB. The spacing between two adjacent such points is fixed to be $\delta$, and hence there are totally $nL/\delta$ such points on the EB. \footnote{For lattice models, the spacing $\delta$ is naturally fixed, and therefore $n$ is finite; for continuous theories, $\delta$ is in principle zero, and $n$ goes to infinity. Technically, when calculating entanglement entropy in continuous theories, one may choose a cutoff and also fix $\delta$ to be finite. We assume $\delta$ is fixed and finite hereinafter.} For subsystem P, the EB anyon states are written as 
\be
\ket{x_1,x_2,\cdots,x_n}_{\textrm{P}},
\ee
where $x_i\in\mathcal{D}$ denotes the boundary anyon on the $i$-th point. 

These states are not norm-$1$; instead, we have
\be\label{Eq:statenorm}
\langle x_1,x_2,\cdots,x_n | x_1,x_2,\cdots,x_n\rangle=\sqrt{d_{x_1}d_{x_2}\cdots d_{x_n}}.
\ee
See Appendix \ref{appd:bdrynorm} for details. The local Hilbert space of the subsystem P is
\be
\mathcal{H}_{\textrm{P}}=\textrm{span}\{\ket{x_1,x_2,\cdots,x_n}_{\textrm{P}}\},
\ee
and also the local Hilbert space for subsystem Q is
\be
\mathcal{H}_{\textrm{Q}}=\textrm{span}\{\ket{\bar x_1, \bar x_2,\cdots,\bar x_n}_{\textrm{Q}}\},
\ee
where $\bar x_i$ is the conjugate anyon of $x_i$. Given the local Hilbert spaces, the ground state of the sphere can then be expressed in terms of the tensor product of the states in $\mathcal{H}_{\textrm{P}}$ and $\mathcal{H}_{\textrm{Q}}$,
\be\label{Eq:statepartition}
\ket{\psi}=\sum_{\{x_i\}} \frac{1}{\sqrt{\Psi}}\ket{x_1,x_2,\cdots,x_n}_{\textrm{P}} \ket{\bar x_1,\bar x_2,\cdots,\bar x_n}_{\textrm{Q}},
\ee
where the summation runs over all allowed configurations of the boundary anyons $\{x_i\}$ and $\Psi$ is a normalization factor to be determined. This definition of tensor product decomposition is artificial and not unique at the model-independent level. In section \ref{subSec:constraint}, we will see that the TEE in this case does not depend on the choice of this definition. As discussed in section \ref{sec:review}, all the boundary anyons should fuse to $1$ due to the constraint on the configurations of the boundary anyons. Therefore, the summation runs over all configurations where the boundary anyons fuse to $1$\footnote{For boundary anyons whose quantum dimension is greater than $1$, there might be multiple fusion channels for these anyons to fuse to $1$. The states of these different fusion channels are orthogonal to each other and hence should be summed individually.}. As such, we can fix the normalization factor to be
\be
\Psi=\sum_{\prod x_i=1}|\langle x_1,x_2,\cdots,x_n | x_1,x_2,\cdots,x_n\rangle|^2=\sum_{\prod x_i=1}d_{x_1}d_{x_2}\cdots d_{x_n}=D^{n-1},
\ee
where $D=\sum_{a\in\mathcal{D}}d_a$. See Appendix \ref{appd:calculate} for a detailed verification of the last equality. We can now partial-trace the subsystem Q and acquire the reduced density matrix of the subsystem P,
\be\label{Eq:rhosph}
\rho_{\textrm{P}}=\sum_{\prod x_i=1}\frac{1}{D^{n-1}} \langle x_1,x_2,\cdots,x_n | x_1,x_2,\cdots,x_n\rangle \ket{x_1,x_2,\cdots,x_n}\bra{x_1,x_2,\cdots,x_n}.
\ee
For an observer in the subsystem P, the classical probability of finding a state $\ket{x_1,x_2,\cdots,x_n}_{\textrm{P}}$ is $\frac{1}{D^{n-1}}d_{x_1}d_{x_2}\cdots d_{x_n}$. The two Hilbert spaces are isomorphic to each other, and verifying the entanglement entropy using either of the reduced density matrix gives equal results. We will omit the subscript P and simply call the Hilbert spaces subsystem Hilbert spaces hereinafter. The entanglement entropy is verified to be
\be
S_{\textrm{EE}}=-\sum_{\prod x_i=1} \frac{d_{x_1}d_{x_2}...d_{x_n}}{D_{n-1}}\textrm{ln}\frac{d_{x_1}d_{x_2}...d_{x_n}}{D_{n-1}}= - n\sum_{a\in\mathcal{D}}\frac{d_a^2}{D}\textrm{ln}\frac{d_a}{D} - \textrm{ln}D.
\ee
See Appendix \ref{appd:calculate} for a detailed verification of the last equality. The term proportional to $n=L/\delta$ is the area-law contribution, and the $-\textrm{ln}D$ is the TEE, which is a direct consequence of the constraint on the boundary anyons. To see this, one can assume that the constraint is absent and repeat the calculation above; the result of entanglement entropy would be the area-law term alone. The constraint rules out some boundary anyon configurations and consequently makes all the points on the EB correlated with each other no matter how far away they are from each other. This correlation is a straightforward demonstration of the \textit{long-range entanglement}. 

One more comment here is that we did not specify any GBC during the analysis. In this case, different GBCs give same TEE, because Eq. \eqref{Eq:corresp} holds for every GBC. We will see that in the cases where the entanglement cuts are open segments, the choice of GBC matters.

\subsection{Case: ground states on a torus}\label{subsec:torus}

\begin{figure}[h!]
    \centering
    \includegraphics[scale=0.45]{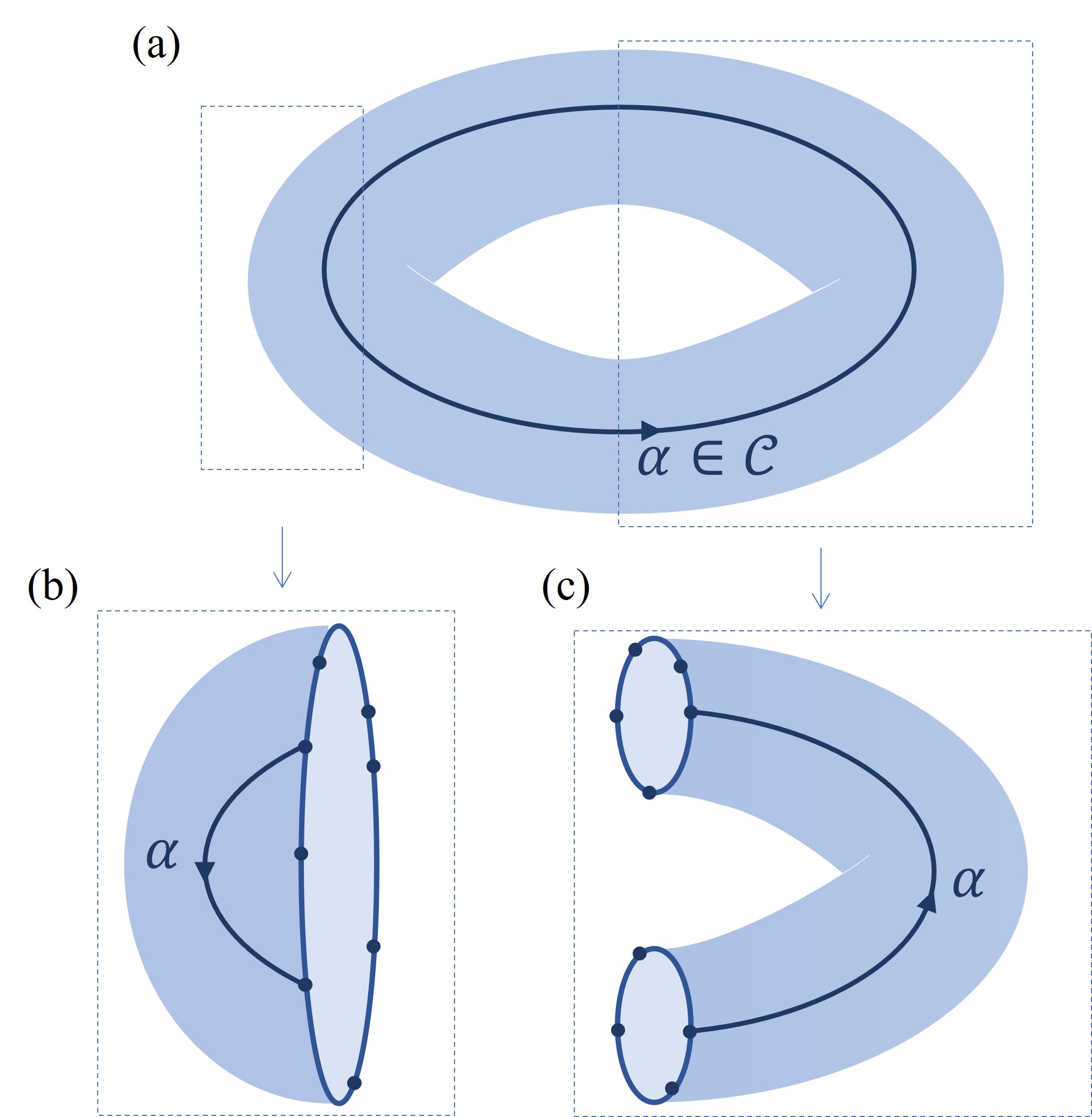}
    \caption{(a) Bi-partitioning a torus. The torus is in the grouns state where a bulk anyon $\alpha$ threads through the torus. (b) The subsystem is still a disk. (c) The subsystem is a cylinder. A bulk anyon $\alpha$ thread through the cylinder.}
    \label{fig5:biparttorus}
\end{figure}

We now consider a specific ground state of a topological order on a torus where an anyon $\alpha$ threads through the torus, as shown in Fig. \ref{fig5:biparttorus}. 

If subsystem P is still a disk (as shown in Fig. \ref{fig5:biparttorus}b), the entanglement entropy is the same as that in the case of a sphere. If the subsystem is a cylinder with two EBs (as shown in Fig. \ref{fig5:biparttorus}c), both EBs will have non-trivial total topological charge. According to the discussion in section \ref{sec:review}, one of the EBs has total charge $\alpha$ and the other has $\bar\alpha$. There is no further correlation between the two EBs. Suppose the two EBs have length $L_1=n_1\delta$ and $L_2=n_2\delta$, and for simplicity, we assume the two EBs have the same GBC. The subsystem Hilbert space is spanned by all the states of boundary anyon configurations that survives the constraint, 
\be\label{Eq:torusHilbert}
\mathcal{H}_{\alpha,\bar\alpha}=\bigoplus_{a,b}\mathcal{H}_{a,b},
\ee
where $\mathcal{H}_{a,b}$ is spanned by the states where the anyons on the two EBs respectively fuse to $a$ and $b$, and $a$ ($b$) runs over all allowed boundary anyons satisfying $W_{\alpha,a}\neq 0$ ($W_{\bar\alpha,b}\neq 0$)\footnote{If $W_{\alpha a}>1$, we should also count the multiplicity of $a$.}. The reduced density matrix is then
\be\label{Eq:torusrho}
\rho_\alpha=\frac{1}{\Psi_\alpha}\sum_{a,b}\sum_{\prod x_i=a,\prod y_i=b}\ket{x_1,x_2,\cdots,x_{n_1};y_1,y_2,\cdots,y_{n_2}}\bra{x_1,x_2,\cdots,x_{n_1};y_1,y_2,\cdots,y_{n_2}},
\ee
where $x_i$ and $y_i$ denotes the anyons respectively on the two EBs, and the first summation runs over all $a,b$ satisfying $W_{\alpha a}\neq 0$ and $W_{\bar\alpha b}\neq 0$. We can then fix the normalization factor,
\begin{align}
{\Psi_\alpha}=&\sum_{a}W_{\alpha,a} \sum_{b}W_{\bar\alpha,b}\sum_{\prod x_i=a}\sum_{\prod y_i=b}\notag\\
&|\langle x_1,x_2,\cdots,x_{n_1} | x_1,x_2,\cdots,x_{n_1}\rangle|^2|\langle y_1,y_2,\cdots,y_{n_2} | y_1,y_2,\cdots,y_{n_2}\rangle|^2\\
=&\sum_{a}W_{\alpha,a} \sum_{b}W_{\bar\alpha,b}\sum_{\prod x_i=a}\sum_{\prod y_i=b} d_{x_1}d_{x_2}\cdots d_{x_{n_1}}d_{y_1}d_{y_2}\cdots d_{y_{n_1}}\\
=&\sum_{a}W_{\alpha,a} \sum_{b}W_{\bar\alpha,b}d_a d_b D^{n_1+n_2-2} = d_{\alpha}d_{\bar\alpha}D^{n_1+n_2-2}=d_{\alpha}^2D^{n_1+n_2-2}.
\end{align}
where the third equality uses the formula introduced in Appendix \ref{appd:calculate} and the forth equality uses Eq. \eqref{Eq:qdim}. The entanglement entropy follows to be, 
\begin{align}
S_{\textrm{EE}}=&-\sum_{a}W_{\alpha,a} \sum_{b}W_{\bar\alpha,b}\sum_{\prod x_i=a}\sum_{\prod y_i=b}\notag\\
&\frac{d_{x_1}d_{x_2}...d_{x_{n_1}}d_{y_1}d_{y_2}...d_{y_{n_1}}}{d_{\alpha}^2D^{n_1+n_2-2}}\textrm{ln}\frac{d_{x_1}d_{x_2}...d_{x_{n_1}}d_{y_1}d_{y_2}...d_{y_{n_1}}}{d_{\alpha}^2D^{n_1+n_2-2}}\\
=&-\sum_{a}W_{\alpha,a} \sum_{b}W_{\bar\alpha,b} \frac{d_ad_b}{d_\alpha^2}(n_1\sum_{x\in\mathcal{D}}\frac{d_x^2}{D}\textrm{ln}\frac{d_x}{D}+n_2\sum_{a\in\mathcal{D}}\frac{d_x^2}{D}\textrm{ln}\frac{d_i}{D}-2\textrm{ln}\frac{D}{d_\alpha})\\
=&- (n_1+n_2)\sum_{x\in\mathcal{D}}\frac{d_x^2}{D}\textrm{ln}\frac{d_x}{D} - 2\textrm{ln}\frac{D}{d_\alpha}.
\end{align}
Again, the term proportional to $n_1+n_2$ is the area-law contribution, and the $-2\textrm{ln}\frac{D}{d_\alpha}$ is TEE. The factor $2$ is due to the two EBs. Each EB contribute to $-\textrm{ln}\frac{D}{d_\alpha}$ due to the constraint on this EB, and the ln$d_\alpha$ term is due to the anyon $\alpha$ that threads through the torus. 

For a generic ground state $\ket{\psi}=\sum_\alpha \psi_\alpha \ket{\psi_\alpha}$ (we abuse the notation of $\psi_k$ to be both the label of the ground state $\ket{\psi_k}$ with an anyon $k$ threads through the torus and its probability amplitude in the generic ground state $\ket{\psi}$.), the entanglement entropy will have an extra term due to the superposition of $\ket{\psi}$. Now, since it is possible to find the total charge of the EBs to be any of the anyons with $\psi_\alpha\neq 0$, the subsystem Hilbert space is then
\be
\mathcal{H}=\bigoplus_\alpha\mathcal{H}_{\alpha,\bar\alpha},
\ee
where $\mathcal{H}_{\alpha,\bar\alpha}$ is the space defined in Eq. \eqref{Eq:torusHilbert} and $\alpha\in\mathcal{C}$ runs over all bulk anyon species with $\psi_\alpha\neq 0$. The two states that respectively belong to $\mathcal{H}_\alpha$ and $\mathcal{H}_\beta$ are orthogonal to each other because the EB total charge is different. The reduced density matrix is then
\be
\rho=\bigoplus_{\alpha} |\psi_\alpha|^2 \rho_\alpha,
\ee
where $\rho_\alpha$ is the density matrix derived in Eq. \eqref{Eq:torusrho}. The TEE follows to be
\be\label{Eq:teetorusgen}
S_{\textrm{TEE}}=-2\sum_{\alpha} |\psi_\alpha|^2 \textrm{ln}\frac{D}{d_\alpha}+\sum_\alpha |\psi_\alpha|^2 \textrm{ln} |\psi_\alpha|^2.
\ee
The $\sum_\alpha |\psi_\alpha|^2 \textrm{ln} |\psi_\alpha|^2$ is the classical von Neumann entropy due to the superposition of the ground basis $\ket{\psi_\alpha}$ and enhances the mixture of the mixed state as expected. For the TEE of a generic state on an arbitrary closed manifold, one can expand the state in terms of the states where each EB have a single total charge. Then, it is straightforward to write down the TEE that composes of three parts: the $-\textrm{ln}D$ term multiplies the number of disconnected EBs, the ln$d_\alpha$ terms of each EB, and the classical term due to the superposition. 

\subsection{The operators of the subsystem}

\begin{figure}[h!]
    \centering
    \includegraphics[scale=0.5]{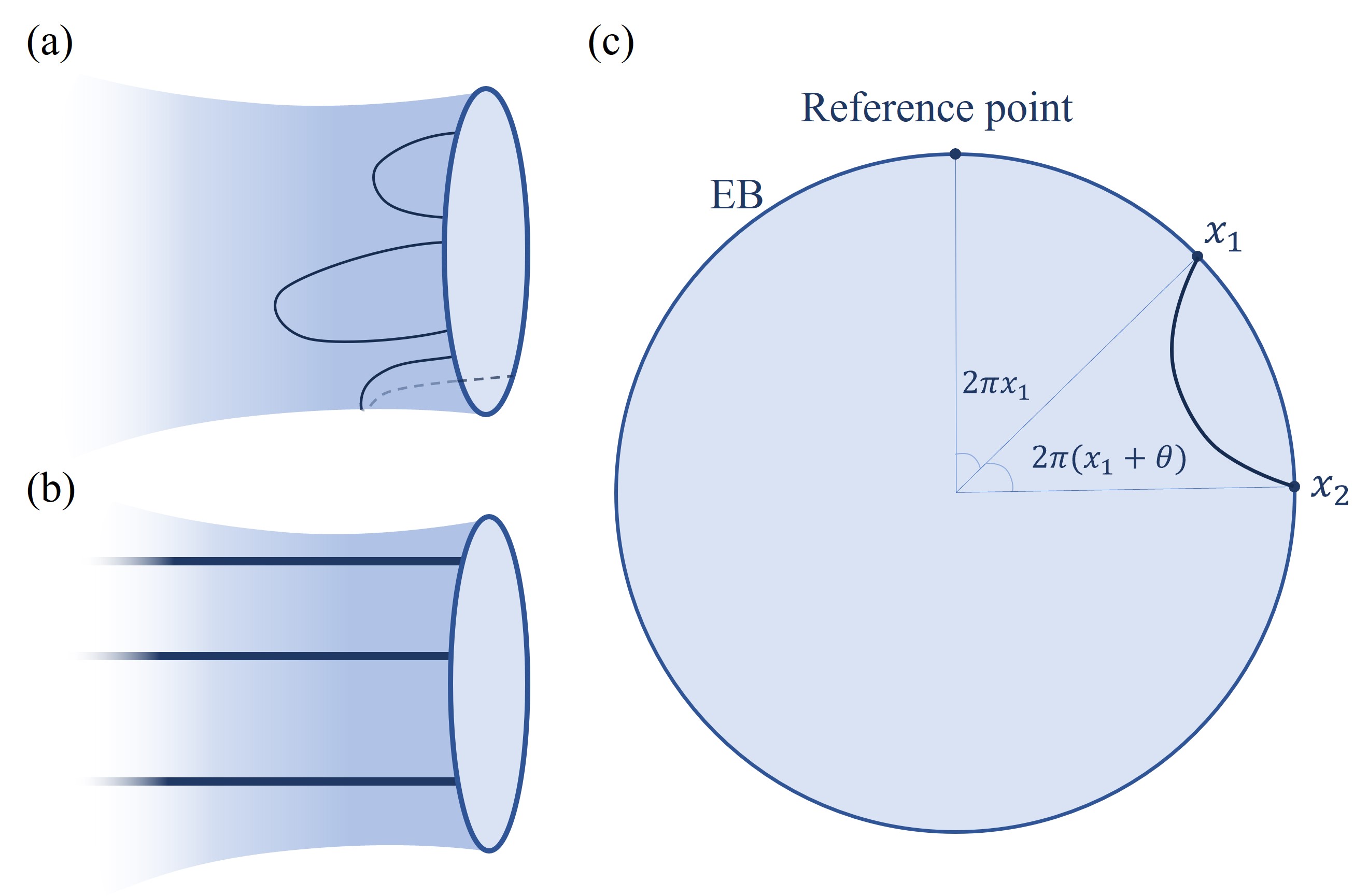}
    \caption{(a) The first type of operators, where each operator has both its endpoints on the same EB. (b) The second type operators, where each operator has only one endpoint on an EB. (c) Choice of the generating set. We choose a reference point on the EB to have $x_0=0$. Other points are labeled by the central angles with respect to the reference point.}
    \label{fig6:opconst}
\end{figure}

As mentioned before, some anyon loops seen by a global observer are cut by the EBs, becoming operators that transform the states in the subsystem Hilbert space. These operators can be classified into two types: the ones with both end points on the same EB, and the ones whose end points are on different EBs. For the second type, the collective behavior of these operators are already captured by the test anyon loops and contribute to the total topological charge of the boundary; for the first type, they create or annihilate boundary defects on the EBs and hence transforms a boundary anyon configuration into another. In fact, for Abelian topological orders, the first type of operators yield to a constraint\footnote{The non-Abelian cases involve non-invertible operators and hence are completely different. We will report the story elsewhere.}. Let us focus on the $D(\mathbb{Z}_2)$ topological order on a sphere, where we only have the first type of operators. There are four kinds of anyon loops respectively correspond to the four kinds of anyons $\{1,e,m,\epsilon\}$. Suppose the EB condenses $e$. Any two points $x_1$ and $x_2$ on the EB are connected by a half loop; each half loop induces two operators: the operator $O_0^{x_1,x_2}$ induced by the half loop of anyons $\{1,e\}$, and a nontrivial operator $O_1^{x_1,x_2}$ induced by the half loop of anyons $\{m,\epsilon\}$. Since the anyons $\{1,e\}$ condense on the EB, the operators $O_0$ do not change the boundary defect configuration and hence are identities in the subsystem Hilbert space. The anyons $\{m,\epsilon\}$ are confined at the EB, and hence the operators $O_1$ creates or annihilates defects on the EB,
\be
O_1^{x_1,x_2} \ket{x_{1},\cdots,x_{2}}=\ket{\tilde x_{1},\cdots,\tilde x_{2}},
\ee
where $\tilde x_i=m^{\{1,e\}}\times x_i$, namely the anyon type on the point $x_i$ is flipped. Since, according to the fusion rule,  
\be
O_1^{x_1,x_2}O_1^{x_2,x_3}=O_1^{x_1,x_3},
\ee
the $O_1$ operators are not independent from each other. We can pick up a generating set by employing the following choice. We arrange the EB into a circle and use an angle $\theta$ to label each point on the cut, as illustrated in Fig. \ref{fig6:opconst}c. A generating set $\mathcal{K}$ consists of the $O_1$ operators that connects two points with fixed angle difference $2\pi\theta$,  
\be
\mathcal{K}_\theta=\{O_1^{x_1,x_2}|x_1-x_2=2\pi\theta\}.
\ee
where $\theta\in(0,1)$ is a fixed irrational number. For each boundary point, there are exactly two operators in the set $\mathcal{K}$ that can create or annihilate an anyon on the point. We fix the $\theta$ to be irrational because otherwise the set cannot generate all the operators.\footnote{In discrete lattice models, it is more convenient to define the generating set as the operators connecting the two adjacent boundary points. In the continuum limit, however, the `adjacent' point of a certain point is not well defined. This is the reason why we need the irrational number $\theta$ here, which makes our choice of the generating set also works in the continuum limit.}. 

All the operators in $\mathcal{K}$ are mutually commute, and hence a state of certain boundary defect configuration can be expressed as
\be\label{Eq:opconfig}
\prod_{x_i}^{\textrm{all points on EB}} (O^{x_i,x_i+\theta})^{k_{x_i}} \ket{0}, 
\ee
where $k_{x_i}=0$ or $1$ for all $x_i$ and $\ket{0}$ is the boundary state with no defect. If, for applicability, we stick to the assumption that there are $n$ points on the EB hosting boundary anyons, then there are $n$ operators in the set $\mathcal{K}$, and the expression \eqref{Eq:opconfig} has $2^n$ different combinations of operators. But this is not the end, the operators in $\mathcal{K}$ yield an extra constraint,
\be
\prod_{x_i}^{\textrm{all points on EB}}O^{x_i,x_i+\theta}=1.
\ee
This constraint is the operator version of the constraint on the EB. The number of independent expressions generated by $\mathcal{K}$ is halved, leaving $2^{n-1}$ independent configurations. The $2^{-1}$ factor finally becomes $S_{\textrm{TEE}}=-\textrm{ln}2$. The construction above is easily generalized to all Abelian topological orders, where we have $k-1$ constraints for a topological order with $k$ species of boundary anyons. 

The operator constraint implies that the boundary anyon fluctuation due to the first type anyon loops should all fuse to $1$, leaving the total charge of the boundary fully determined by the second type anyon loops. In this sense, we can separately take the $-$ln$D$ term and the ln$d_\alpha$ term as the contribution of the first type anyon loops and the second type anyon loops respectively. 

\subsection{Local vs. Topological}\label{subSec:constraint}

The operator constraint serves as a good criterion to distinguish TEE from locally originated contributions. In particular, the tensor product decomposition in Eq. \eqref{Eq:statepartition} is not unique at the model-independent level. One can use other definition of the decomposition, which will lead to off-diagonal terms in the reduced density matrix Eq. \eqref{Eq:rhosph}. Such off-diagonal terms make some points on the EB coherent with each other and will contribute to the entanglement entropy. If the coherence violates the homogeneity of the EB, the contribution will be a sub-leading term. For example, for the ground state of a $D(\mathbb{Z}_2)$ on a sphere, we can pick up two points $x_p$ and $x_q$ on the EB and make them coherent manually. The reduced density matrix will have extra off-diagonal terms between the states $\ket{x_1,\cdots,x_p,\cdots,x_q,\cdots,x_n}$ and $\ket{x_1,\cdots,\tilde x_p,\cdots,\tilde x_q,\cdots,x_n}$. Then, there will be a $\frac{1+\lambda}{2}\textrm{ln}\frac{1+\lambda}{2}+\frac{1-\lambda}{2} \textrm{ln}\frac{1-\lambda}{2}$ term in the entanglement entropy, with $\lambda$ quantifying the strength of the coherence (See Appendix \ref{appd:localterm} for a detailed derivation of this term.). This kind of terms due to inhomogeneity of the EB is neither related to the EB constraint nor topologically invariant. In the effective theories, such terms are naturally circumvented because the EBs are chosen to be homogeneous. At model-independent level, there is a more fundamental way to distinguish these terms from TEE: by checking the operator constraint. We add or remove coherence in the density matrix while keeping the operator constraint unchanged, until we reach a homogeneous EB. Then, the sub-leading term derived from the homogeneous EB is TEE. For the construction we introduced in this section, namely the EBs are closed loops and have fixed GBCs, the operator constraint is always independent of the coherence. 

We require the EBs to be homogeneous in the the rest of this paper, such that the locally originated sub-leading terms are avoided. 

\section{The boundary defect picture: open EBs}

In this section, we investigate the cases where the EBs touches gapped boundaries. In this section, we adopt the simplest EB condition where each EB have a fixed GBC. Unlike the cases of closed entanglement cuts, different GBCs on the EB will give different TEE. This ambiguity happens because the anyon species that can be hosted at the junction varies when changing the GBC. 

\subsection{Case: ground state on a disk}\label{subSec:disk}

\begin{figure}[h!]
    \centering
    \includegraphics[scale=0.45]{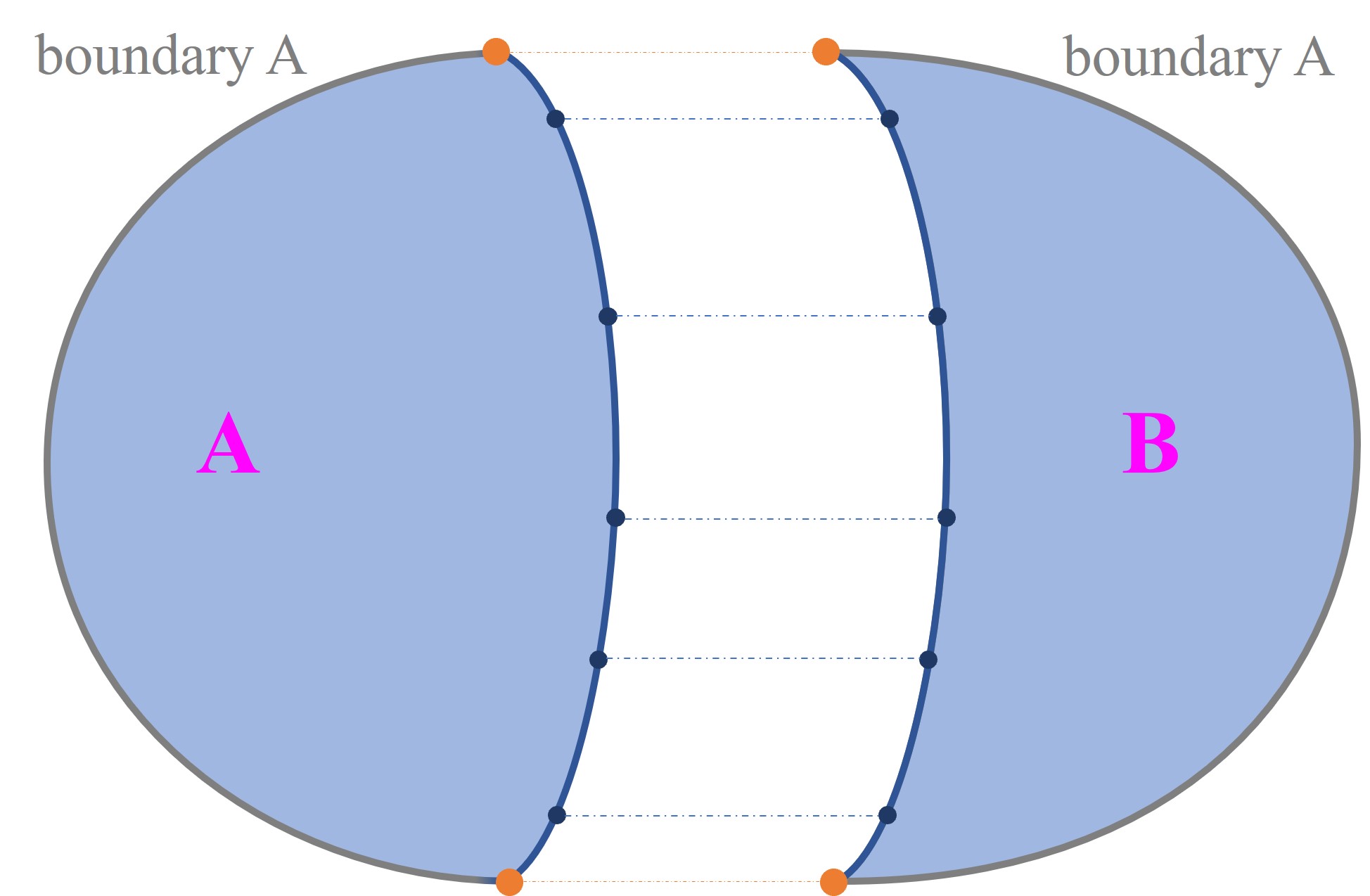}
    \caption{Bi-partitioning a disk. We suppose there are $n$ discrete points on the EB segment (excluding the two junctions) that can host boundary anyons.}
    \label{fig7:bipartdisk}
\end{figure}

We first consider the simplest case: the ground state of a topological order on a disk. Each subsystem is a disk with two boundary segments, as shown in Fig. \ref{fig7:bipartdisk}. We suppose that the GBCs of the physical boundary and the EB are $A$ and $B$. While there is no anyon on the physical boundary segment, there can be anyons on the EB and the two junctions. The subsystem Hilbert space $\mathcal{H}^{A|B}$ is then spanned by
\be
\ket{x_1,x_2,\cdots,x_n;z_1,z_2},
\ee
where $x_i\in\mathcal{D}^B$ represents for the anyon on the $i$-th point on the EB, and $z_i\in\mathcal{D}^{A|B}$ labels the anyon on the $i$-th junction. According to the discussions in section \ref{sec:review}, one can measure the total charge of the EB by a half loop near the EB, and the result in this case would be a trivial total charge $1^A$. Therefore, the constraint on the boundary defects is
\be
z_1\times x_1\times \cdots \times x_n\times z_2 = 1^A.
\ee

The reduced density matrix follows to be
\begin{align}\label{Eq:disk1m}
\rho^{B=\{1,m\}}=\frac{1}{\Psi}\sum_{z_1(\prod x_i)z_2=1^A} d_{z_1}d_{x_1}\cdots d_{x_n} d_{z_2}\ket{x_1,x_2,\cdots,x_n;z_1,z_2}\bra{x_1,x_2,\cdots,x_n;z_1,z_2}.
\end{align}

The TEE depend on the choice of the GBC of the EB. To be concrete, we focus on the $D(\mathbb{Z}_2)$ topological order. Suppose both the physical boundary and the EB have GBC $A=B=\{1,m\}$. Then, $\mathcal{H}^{A|B}$ is $2^{n+1}$ dimensional, and it is straightforward to find that $\Psi=2^{n+1}$. The entanglement entropy follows to be
\be
S_{\textrm{EE}}^{B=\{1,m\}}=-\sum_{z_1(\prod x_i)z_2=1^A} \frac{d_{z_1}d_{x_1}...d_{x_n}d_{z_2}}{2^{n+1}}\textrm{ln}\frac{d_{z_1}d_{x_1}...d_{x_n}d_{z_2}}{2^{n+1}}= - (n+1)\textrm{ln}\frac{1}{2}.
\ee
On the other hand, if the EB have GBC $B=\{1,e\}$, then $\mathcal{H}^{A|B}=$span$\{\ket{x_1,\cdots,x_n;\sigma,\sigma}\}$ is $2^n$ dimensional. The reduced density matrix is then
\be\label{Eq:disk1e}
\rho^{B=\{1,e\}}=\frac{1}{\Psi}\sum_{\{x_i\}} d_{\sigma}d_{x_1}\cdots d_{x_n} d_{\sigma}\ket{x_1,x_2,\cdots,x_n;\sigma,\sigma}\bra{x_1,x_2,\cdots,x_n;\sigma,\sigma}.
\ee
The summation runs over all configurations of $\{x_i\}$ because $\sigma\times (\prod x_i)\times\sigma=1^A+e^A$ automatically satisfies the constraint. Again, $\Psi=2^{n+1}$, and the entanglement entropy is
\be
S_{\textrm{EE}}^{B=\{1,e\}}=-\sum_{\{x_i\}} \frac{\sqrt{2}d_{x_1}...d_{x_n}\sqrt{2}}{2^{n+1}}\textrm{ln}\frac{\sqrt{2}d_{x_1}...d_{x_n}\sqrt{2}}{2^{n+1}}= - n\textrm{ln}\frac{1}{2}.
\ee
Immediately, one wonders which term is the area-law contribution, and which term is TEE? Note that we assumed there are $n$ points on the boundary that can host boundary anyons, and the spacing between two adjacent points is $\delta$. When the EB is open, we have $L=(n+1)\delta$, and hence the area-law term should be proportional to $(n+1)$. We therefore recognize that $S_{\textrm{TEE}}^{B=\{1,m\}}=0$ and $S_{\textrm{TEE}}^{B=\{1,e\}}=-$ln$2$. 

It is worth mentioning that in ref \cite{chen_entanglement_2018}, the authors used the QD model to calculate the TEE of this case, and the area-law term is taken to be $(n+2)$ln$2$. In the QD model, the entanglement cut can be viewed as a gapped boundary with `charge' excitations (excitations reside on vertices) on it, which is the $e$ anyon in $D(\mathbb{Z}_2)$. Therefore, calculating the entanglement entropy using the QD models is equivalent to employing the EB condition $B=\{1,m\}$ in our picture. The $(n+2)$ factor takes the two junction points on an equal footing with other points on the EB. Nevertheless, the two junctions are naturally different from other points; choosing the $(n+2)$ln$2$ term to be area-law is not proper. We will see in section \ref{sec:fold} that a homogeneous open EB should have area-law term proportional to $(n+1)$.

\begin{figure}[h!]
    \centering
    \includegraphics[scale=0.45]{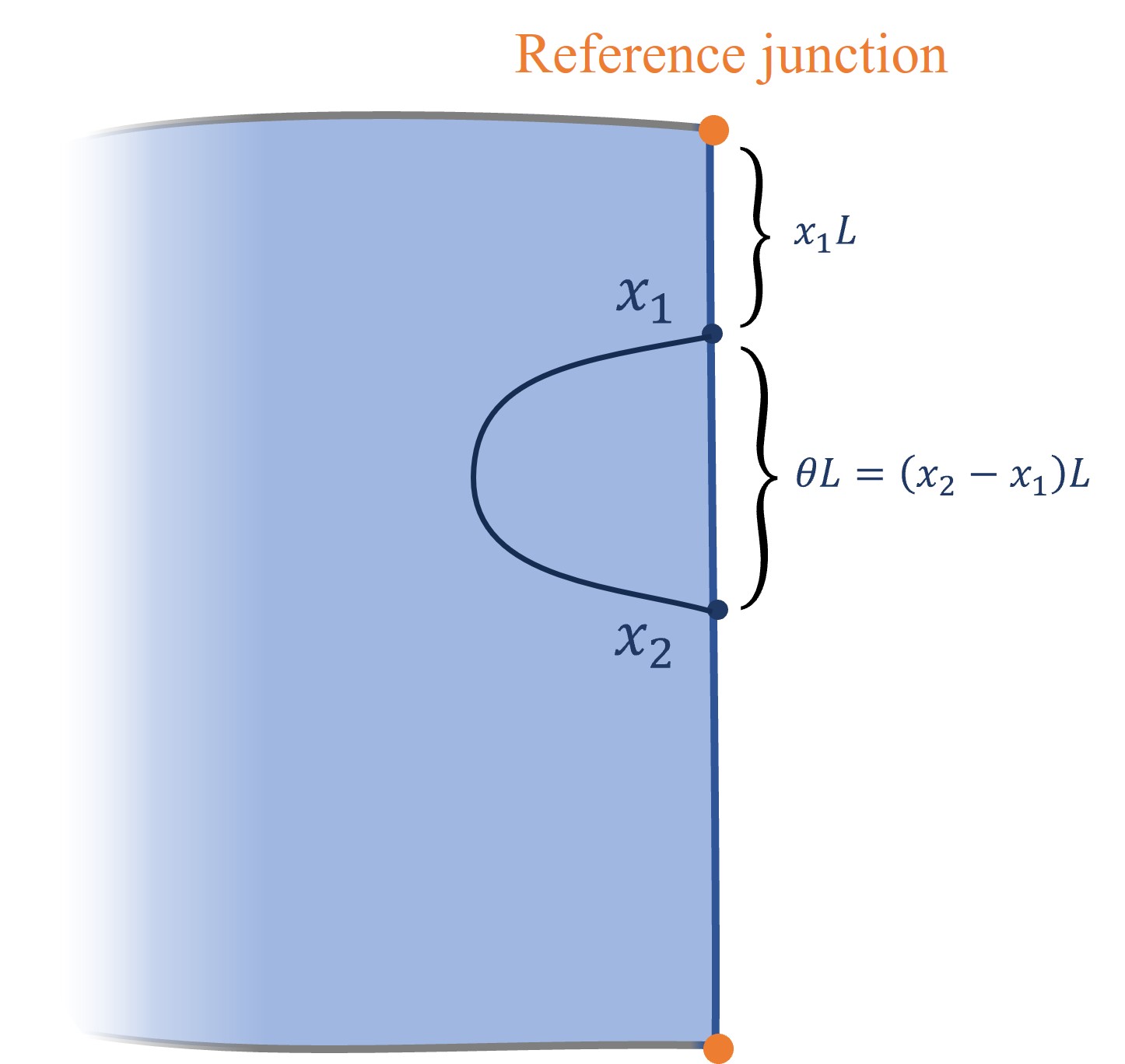}
    \caption{The convention of labelling the points on the EB. The generating set consists of all operators with fixed $x_j-x_i$.}
    \label{fig8:opconsopen}
\end{figure}

Obviously, the TEE depends on the choice of EB condition because the junction anyon species varies when employing different GBCs. This dependence is also shown in the operator constraint. For the open cut cases, we choose the generating set by employing the following convention. The EB is an open segment with length $L$; a point on the EB can be represented by a number $x_i\in[0,1]$, such that the distance between the point and $z_1$ is $x_i L$, as shown in Fig. \ref{fig8:opconsopen}. We choose the generating set of the operators to be all the operators that connects two points whose distance is $\theta$
\begin{align}
\{O^{x_i,x_j}|x_j=\left\{\begin{array}{ll} (x_i+\theta) L & \textrm{if $(x_i+\theta) L < L$ } \\ (x_i+\theta-1) L & \textrm{if $(x_i+\theta) L > L$}\end{array}\right.\},
\end{align}
where $\theta$ is again an irrational number. As such, the number of the operators in the generating set equals the number of boundary points as before. Multiplying these operators together results in, instead of the identity in the cases of closed cuts, the operator that connects the two junctions,
\begin{equation}\label{Eq:opencons}
\prod_{x_i} O^{x_i,x_i+\theta}=O^{z_1,z_2}.
\end{equation}
The dependence on the EB condition is encoded in $O^{z_1,z_2}$, which transforms a junction anyon configuration into another. If the subsystem Hilbert space is $\mathcal{H}^{\{1,e\}|\{1,m\}}$, the operator $O_{z_1,z_2}$ is again an identity matrix, because there is only one type of junction anyon allowed; for $\mathcal{H}^{\{1,m\}|\{1,m\}}$, however, $O_{z_1,z_2}$ is no longer an identity,
\be
O_{z_1,z_2}:\hspace{5pt} \ket{x_1,\cdots,x_n;z_1,z_2}\leftrightarrow\ket{x_1,\cdots,x_n;\tilde z_1,\tilde z_2}.
\ee
The operator constraint also varies when changing the EB condition. A direct observation is that adding coherence between the two junction points will modify the explicit form of the operator $O_{z_1,z_2}$ and accordingly the value of TEE.

It is similar to derive the TEE for the case where $A=\{1,e\}$. We summarize the results in the following table.

\begin{table}[H]
\centering
\begin{tabular}{|c|c|c|}
    \hline
    $S_{\textrm{TEE}}$ & A=$\{1,e\}$ & A=$\{1,m\}$\\
    \hline
    B=$\{1,e\}$ & 0 & $-$ln$2$ \\
    \hline
    B=$\{1,m\}$ & $-$ln$2$ & 0\\
    \hline
\end{tabular}
\end{table}

For a fixed EB condition, different GBCs on the physical boundary also gives different TEE. In contrast, the TEE is $-\frac{1}{2}$ln$2$ regardless of the physical GBC according to the calculation in the $K$-matrix theory. This mismatch indicates that neither of the EB conditions we used in this section is compatible with the EB condition in the $K$-matrix theory. We will treat this mismatch in Section \ref{sec:fold} by introducing a general construction of the subsystem density matrix.

\subsection{Case: ground states on a cylinder}

\begin{figure}[h!]
    \centering
    \includegraphics[scale=0.4]{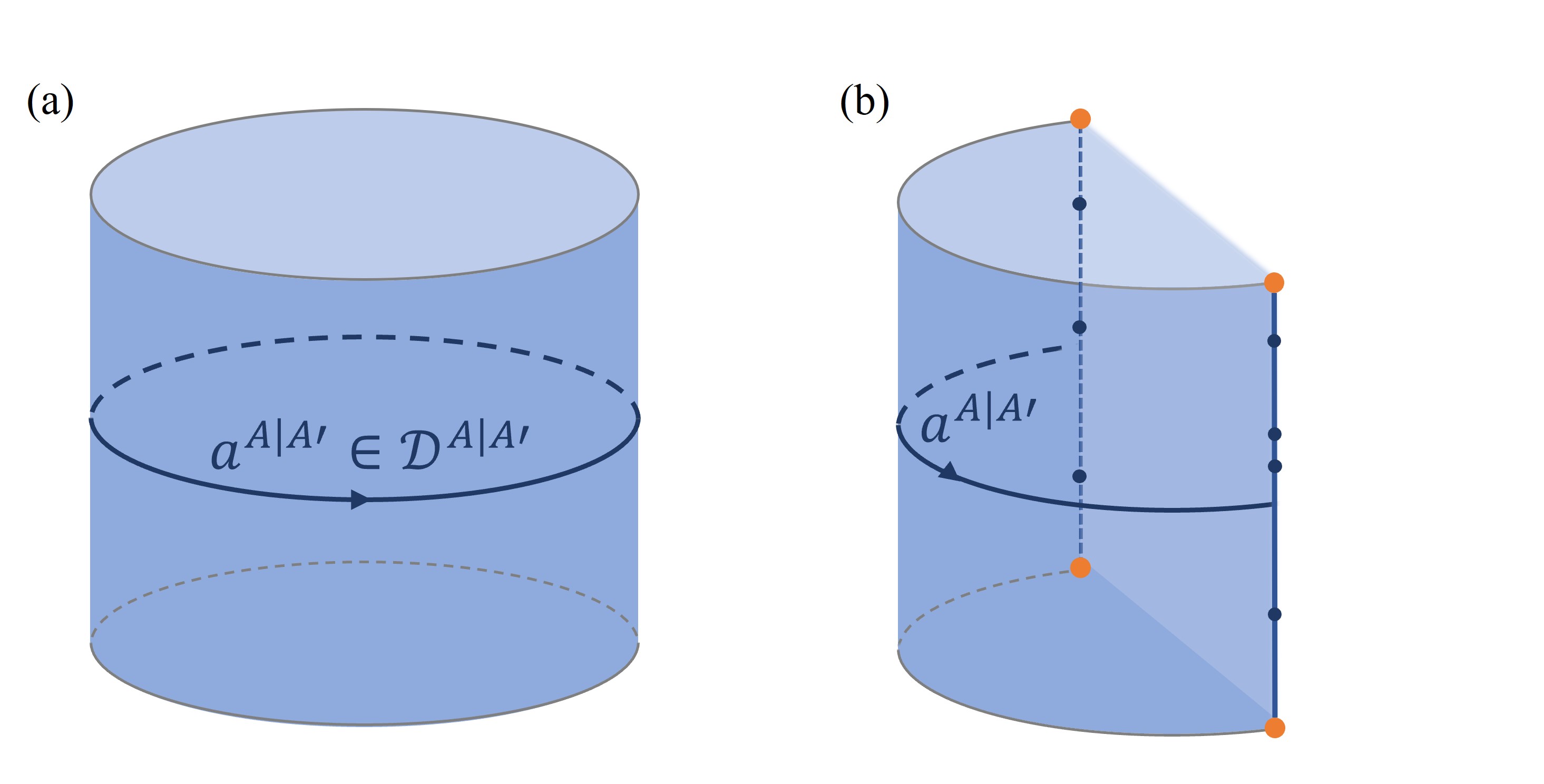}
    \caption{(a) A cylinder is in the ground state labeled by a confined anyon $a^{A|A'}$. (b) Bi-partitioning a cylinder.}
    \label{fig9:bipartcyl}
\end{figure}

Before introducing the general construction, we would like to first consider the ground states on a cylinder. The cylinder cases are analogous to the torus cases, in the sense that the total charge of an EB can be a non-trivial anyon. Suppose the cylinder has GBCs $A$ and $A'$. We focus on the specific ground state that correspond to a confined anyon $\ket{\psi_{a^{A|A'}}}$, such that the total charge of both EBs is $a^{A|A'}\in\mathcal{D}^{A|A'}$. We further suppose the EBs has GBC $B$ and $B'$, and there are $n$ ($m$) anyon points on the boundary $B$ ($B'$), as shown in Fig. (\ref{fig9:bipartcyl}). 

The subsystem Hilbert space is then 
\be
\mathcal{H}^{A,A'|B,B'}=\textrm{span}\{\ket{x_1,\cdots,x_n;z_1,z_2;y_1,\cdots,y_m;z_3,z_4}\},
\ee
with the constraints
\begin{align}\label{Eq:cylincons}
z_1^{A|B}\times(\prod_i x_i^{B})\times z_2^{B|A'}=a^{A|A'};\hspace{5pt}z_3^{A|B'}\times(\prod_i y_i^{B'})\times z_4^{B'|A'}=a^{A|A'},
\end{align}
where $x_i$'s denote the anyons on the EB $B$, $y_i$'s denote the anyons on the EB $B'$, and $z_i$'s respectively denote the anyons on the four junction points. The reduced density matrix is then
\be
\rho_{a^{A|A'}}=\sum_{\{x_i,y_i,z_i\}} \prod d_{z_i} \prod d_{x_i} \prod d_{y_i} \ket{x_1,\cdots;z_1,z_2;y_1,\cdots;z_3,z_4}\bra{x_1,\cdots;z_1,z_2;y_1,\cdots;z_3,z_4},
\ee
where the summation runs over all boundary anyon configurations satisfying the constraint Eq. \eqref{Eq:cylincons}. The resultant TEE is $-\textrm{ln}D^B-\textrm{ln}D^{B'}+2\textrm{ln}d_{a^{A|A'}}$. The ln$D^B$ and ln$D^{B'}$ terms are analogous to the ln$D$ term in the closed EB cases, but now they depend on the choice of EB conditions. The $\textrm{ln}d_{a^{A|A'}}$ term is analogous to the ln$d_{\alpha}$ term in the closed EB cases because they both attribute to the total charge of the EBs. A generic cylinder ground state can be expanded in terms of the basis states $\ket{\psi_{a^{A|A'}}}$, and the probability amplitude of each basis state will again result in a classical part $\sum_{a^{A|A'}}|\psi_{a^{A|A'}}|^2$ln$|\psi_{a^{A|A'}}|^2$, analogous to the classical part in Eq. \eqref{Eq:teetorusgen}.

%the difference is that in the closed EB cases, the total charge of an EB is a bulk anyon $\alpha$, while in open EB cases, the total charge of an EB is a boundary defect $a$. 

\section{A general state decomposition for $D(\mathbb{Z}_2)$}\label{sec:fold}

As mentioned in before, TEE calculated in the $K$-matrix theory and the lattice models does not match. In this section, we focus on the $D(\mathbb{Z}_2)$ topological order and introduce a general construction of the state decomposition that addresses the ambiguity. As mentioned in Section \ref{subSec:constraint}, defining the decomposition amounts to constructing subsystem density matrix. We construct a general form of the reduced density matrix that can fully characterize the ambiguity.

\subsection{The EM duality in the folding trick}
%For the simplest example -- the $D(\mathbb{Z}_2)$ topological order on a disk, results calculated in the $K$-matrix theory, which employs the \textit{folding trick}, shows that the TEE is $\frac{1}{2}$ln$2$ regardless of the GBC of the physical boundary. This result has manifest EM duality, i.e. changing the role of $e$ and $m$ in the theory leads to no difference in the quantity, while the constructions in the above discussions assumed that the EB condition is a single GBC for each EB, which does not respect the EM duality. reproduce the result calculated in the $K$-matrix theory by choosing proper a EB condition and adding specific coherence terms.

\begin{figure}[h!]
    \centering
    \includegraphics[scale=0.4]{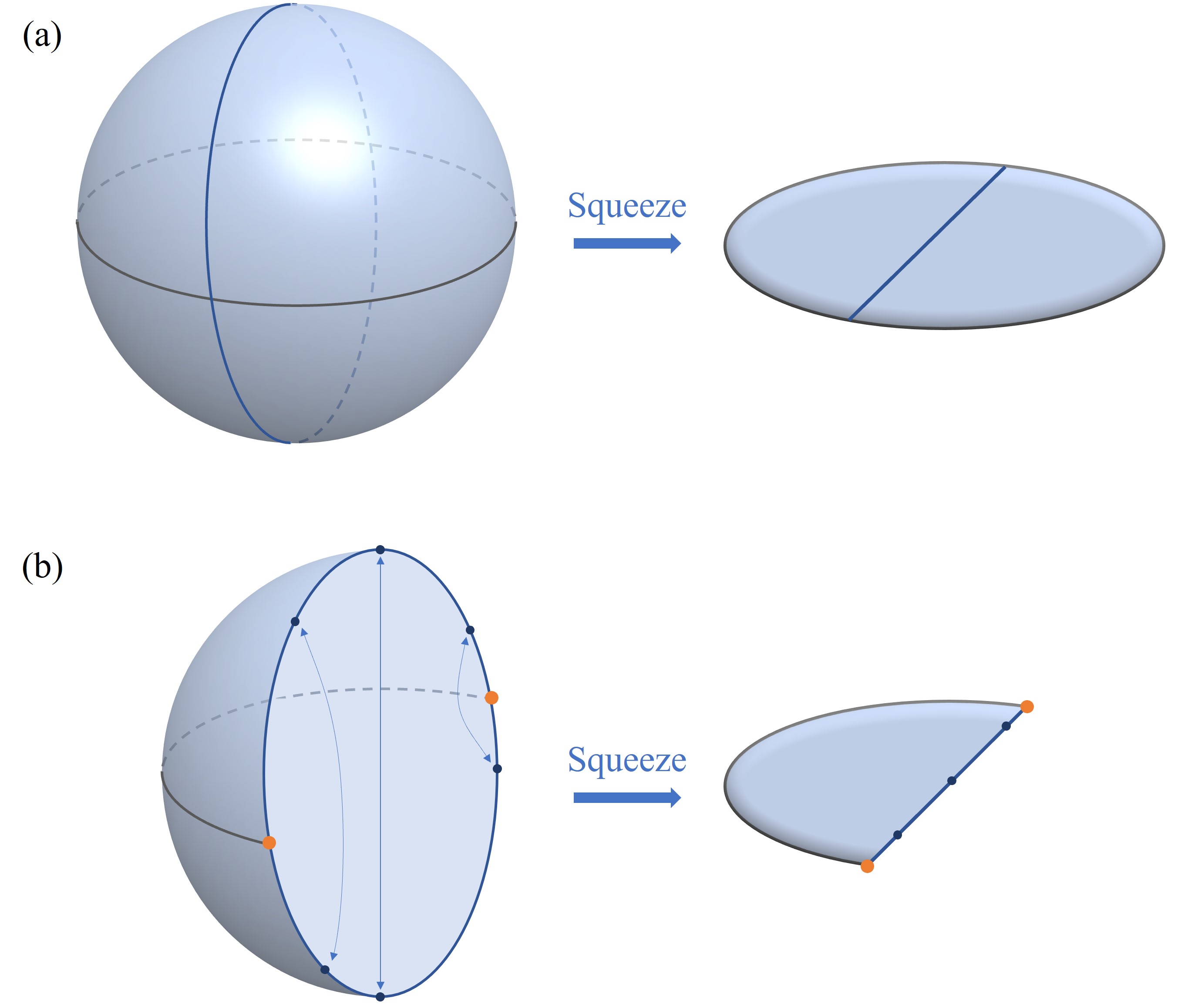}
    \caption{(a) A chiral $\mathbb{Z}_2$ topological order on a sphere (left) is squeezed into a $D(\mathbb{Z}_2)$ topological order on a disk. The equator on the sphere becomes the gapped boundary of the disk. The blue large circle is the entanglement cut, which is also folded into an open line segment on the disk. (b) The special points on the EB of the sphere are paired and identified on the EB of the disk, except that the two orange points that become the junctions on the disk. }
    \label{fig10:foldtrick}
\end{figure}

The key idea of calculating TEE in the $K$-matrix theory is the folding trick. In ref \cite{shen_ishibashi_2019}, the authors constructed the Ishibashi state that describes the entanglement cut via unfolding the non-chiral topological order. For example, for the $D(\mathbb{Z}_2)$ on a disk, one can unfold the disk into chiral $\mathbb{Z}_2$ topological order on a sphere, as shown in the Fig. \ref{fig10:foldtrick}. The chiral $\mathbb{Z}_2$ anyon on the upper and the lower layer of the squashed sphere can be respectively thought as the $e$ and $m$ anyons of the $D(\mathbb{Z}_2)$ in the bulk, and the $\epsilon$ anyon is the composition of the $e$ and $m$. 

Given the picture of the folding trick, we can understand why the area-law term should be proportional to $n+1$. Suppose there are $n$ points that can host boundary anyons on the EB of the disk, if we unfold the half disk into a semi-sphere, the anyon points will split into $2n$ special points\footnote{The unfolded EB is usually a gapless boundary described a CFT state, such that there may not be boundary anyons on the unfolded EB. Therefore we simply call these unfolded points `special points'.}. Taking the two junctions into consideration, there are in all $(2n+2)=\frac{2L}{\delta}$ special points on the unfolded EB. The homogeneity of the EB then means that these $2n+2$ special points are on an equal footing, and hence the area-law term should be proportional to $2n+2$. 

If one regard the systems before and after the folding trick to be completely equivalent, one can directly find the properties of the folded system from that of the unfolded system. For example, since we already know that the TEE of chiral $\mathbb{Z}_2$ topological order on a sphere is ln$\sqrt{2}$, one may think the TEE of $D(\mathbb{Z}_2)$ on a disk is also ln$\sqrt{2}$. This folding process indicates that the EB condition in the $K$-matrix theory respects EM duality: the EB treats equally the $\mathbb{Z}_2$ anyon on the upper layer and the lower layer, which are respectively $e$ and $m$ in $D(\mathbb{Z}_2)$. 

%For the simplest example -- the $D(\mathbb{Z}_2)$ topological order on a disk, results calculated in the $K$-matrix theory, which employs the \textit{folding trick}, shows that the TEE is $\frac{1}{2}$ln$2$ regardless of the GBC of the physical boundary. This result has manifest EM duality, i.e. changing the role of $e$ and $m$ in the theory leads to the same result. 

\subsection{The general EB condition}

The EM duality in the folding trick hints us to extend the EB condition to include both GBCs of the $D(\mathbb{Z}_2)$ topological order. If an EB have a single GBC $\{1,e\}$ ($\{1,m\}$) as its EB condition, we use an extra superscript $^e$ ($^m$) to label the defect configuration states $\ket{\{x_i\}}^e$ ($\ket{\{x_i\}}^m$) on this EB. The Hilbert space spanned by $\ket{\{x_i\}}^e$ ($\ket{\{x_i\}}^m$) is denoted as $\mathcal{H}^e$ ($\mathcal{H}^m$). The general EB condition for a certain EB induces a composite subsystem Hilbert space,
\be
\mathcal{H}^{\textrm{EM}}=\mathcal{H}^e\oplus\mathcal{H}^m.
\ee
Namely, the general Hilbert space collects the states with different GBCs on the EB. Let us focus on the $D(\mathbb{Z}_2)$ on a disk with the physical boundary having GBC $A=\{1,e\}$. We have $\mathcal{H}^e=\mathcal{H}^{\{1,e\}|\{1,e\}}$ and $\mathcal{H}^m=\mathcal{H}^{\{1,e\}|\{1,m\}}$, where $\mathcal{H}^{\{1,e\}|\{1,e\}}$ is spanned by $\ket{\{x_i\};z_1,z_2}^e$, and $\mathcal{H}^{\{1,e\}|\{1,m\}}$ is spanned by $\ket{\{x_i\};\sigma,\sigma}^m$, with the superscript denoting which Hilbert space the states belong to. A naive choice of the reduced density matrix is a direct sum of the two reduced density matrix in Eq. \eqref{Eq:disk1e} and Eq. \eqref{Eq:disk1e},
\begin{align}
\rho^{p}=&p \rho^e \otimes (1-p)\rho^m\\
=&p\sum_{z_1(\prod x_i)z_2=1}\frac{d_{z_1}d_{x_1}...d_{x_n}d_{z_2}}{2^{n+1}}\ket{x_1,\cdots,x_n;z_1,z_2}\bra{x_1,\cdots,x_n;z_1,z_2}+\\
&(1-p)\sum_{\{x_i\}}\frac{d_{\sigma}d_{x_1}...d_{x_n}d_{\sigma}}{2^{n+1}}\ket{x_1,\cdots,x_n;\sigma,\sigma}\bra{x_1,\cdots,x_n;\sigma,\sigma},
\end{align}
where $p\in[0,1]$ is a parameter that cannot be model-independently determined. The parameter $p$ is the classical probability of finding the EB state in $\mathcal{H}^e$. For the lattice models, we have $p=0$, such that the charges are confined and fluctuate on the EB. For the folding trick, we should have $p=\frac{1}{2}$, such that the probability of finding the EB state in $\mathcal{H}^e$ and $\mathcal{H}^m$ are equal. This equality is the basic requirement of EM duality: the EB has $50\%$ chance condensing $e$ and $50\%$ chance condensing $m$. This basic requirement is a necessary condition for the $e$ anyons and the $m$ anyons being treated equally on the EB. 

Nevertheless, this naive choice of the reduced density matrix $\rho^{p}$ can not reproduce the result calculated in the $K$-matrix theory. The entanglement entropy is verified to be
\begin{align}
S_{\textrm{EE}}=&\sum_{z_1(\prod x_i)z_2=1}\frac{pd_{z_1}d_{x_1}...d_{x_n}d_{z_2}}{2^{n+1}}\textrm{ln}\frac{pd_{z_1}d_{x_1}...d_{x_n}d_{z_2}}{2^{n+1}}\\
&-\sum_{\{x_i\}}\frac{(1-p)d_{\sigma}d_{x_1}...d_{x_n}d_{\sigma}}{2^{n+1}}\textrm{ln}\frac{(1-p)d_{\sigma}d_{x_1}...d_{x_n}d_{\sigma}}{2^{n+1}}\\
=&(n+1)\textrm{ln}2-p\textrm{ln}p-(1-p)\textrm{ln}(2-2p).
\end{align}

For $p=\frac{1}{2}$, the sub-leading term is positive. If we look at the operators of this subsystem, we will find two independent sets of operators $O_e$ and $O_m$ that respectively transform the states within $\mathcal{H}^e$ or $\mathcal{H}^m$; each set has its own operator constraint, and no operator can transform states between $\mathcal{H}^e$ and $\mathcal{H}^m$. This doubled operator constraints tell us that the whole Hilbert space $\mathcal{H}^{\textrm{EM}}$ is too large. To reproduce the results calculated in the $K$-matrix theory, the density matrix $\rho^{p=\frac{1}{2}}$ should have have some zero eigenvalues, such that not all the states in $\mathcal{H}^{\textrm{EM}}$ contribute to entanglement entropy. We now construct such a mixed state.

\subsection{The general density matrix}

Technically, we can still work in the basis of $\ket{x_1,\cdots,x_n;z_1,z_2}$ and $\ket{x_1,\cdots,x_n;\sigma,\sigma}$ and add coherence in the density matrix $\rho^{p}$. According to the discussion in section \ref{subSec:constraint}, adding coherence may result in local contributions to TEE. How to prevent from the inhomogeneity while adding coherence? 

The answer lies in the exact EM duality of the topological orders with gapped boundaries \cite{wang_electricmagnetic_2020}. Mathematically, there is a map between the allowed set of anyons on boundaries with different GBCs induced by the EM duality. For the $D(\mathbb{Z}_2)$ topological order, this map is an isomorphism, namely $\mathcal{D}^{\{1,e\}}\cong \mathcal{D}^{\{1,m\}}$. This isomorphism induces a natural map between the two Hilbert spaces $\mathcal{H}^e$ and $\mathcal{H}^m$. Let us first see how this map work in the case of closed EB.

\subsubsection{Closed EB}

Let us go back to the case of $D(\mathbb{Z}_2)$ on a sphere, where $\mathcal{H}^e$ and $\mathcal{H}^m$ are isomorphic to each other. It is simple to establish the map
\be\label{Eq:closemap}
\ket{x_1,x_2,\cdots,x_n}^e \leftrightarrow \ket{x_1,x_2,\cdots,x_n}^m.
\ee
With this map, the basis states of the Hilbert space $\mathcal{H}^{\textrm{EM}}$ can be decomposed into $2^{n-1}$ pairs, and the reduced density matrix can be rearranged as
\be\label{Eq:emdualsph}
\rho^p=\frac{1}{\Psi}\underbrace{\bordermatrix{
&\ket{\{x_i\}}^e&\ket{\{x_i\}}^m\cr
&p&0\cr
&0&1-p}\oplus\cdots}_{2^{n-1}\textrm{ blocks}}
\ee
A $\ket{\{x_i\}}$ stands for a certain defect configuration on the EB. According to the discussions above, there are $2^{n-1}$ different configurations; each of the $2^{n-1}$ blocks corresponds one configuration. Adding coherence terms within each block in Eq. \eqref{Eq:emdualsph} but not between the blocks prevents us from the local contributions. For the EB being homogeneous, each block should receive the same coherence term. Therefore, the general form of the reduced density matrix for $\mathbb{Z}_2$ topological order in closed EB cases is
\be\label{Eq:generalclosed}
\rho^p=\frac{1}{\Psi}\underbrace{\bordermatrix{
&\ket{\{x_i\}}^e&\ket{\{x_i\}}^m\cr
&p&c\cr
&c^*&1-p}\oplus\cdots}_{2^{n-1}\textrm{ blocks}},
\ee
where $c$ is the coherence term to be added. To reproduce the results compatible with the folding trick, at least one of the eigenvalues of the block should become zero, such that the dimension of the physical space is reduced. As such, the only possible coherence term, in this simple case, is then
\be
c=\sqrt{p(1-p)}e^{i\phi},
\ee
where $\phi$ is a parameter that cannot be determined model-independently. Diagonalizing this new density matrix gives
\be
\rho=\frac{1}{\Psi}\underbrace{\bordermatrix{
&\ket{\phi^+,\{x_i\}}&\ket{\phi^-,\{x_i\}}\cr
&1&0\cr
&0&0}\oplus\cdots}_{2^{n-1}\textrm{ blocks}}
\ee
and the $S_{\textrm{TEE}}=-$ln$2$ is independent of $p$. The eigenvectors of this density matrix are $\ket{\phi^{\pm},\{x_i\}}=\sqrt{(1-p)}\ket{\{x_i\}}^e\pm \sqrt{p}e^{i\phi/2}\ket{\{x_i\}}^m$, where the vectors with plus (minus) sign has eigenvalue $1$ ($0$). In the perspective of a local observer in the subsystem, what she sees is a classical ensemble of these $\ket{\phi^{+},\{x_i\}}$ states. If we require the EM duality, namely $p=\frac{1}{2}$, each $\ket{\phi^{+},\{x_i\}}$ is a superposition of a $\mathcal{H}^e$ state and a $\mathcal{H}^m$ state with equal probability. This equality is the advanced requirement of EM duality: every states that a subsystem observer can see treat $e$ and $m$ equally. In this simple case, this stronger requirement is recovered so long as the basic requirement is fulfilled. 

What happens to the operator constraint? In fact, the choice of the map is not unique. Imposing any fixed combinations of $O_e$ ($O_m$) operators on the left (right) hand side in expression \eqref{Eq:closemap} also gives valid maps. Using different maps merely changes the explicit expression of $\ket{\phi^{\pm}}$ but results in no physical consequence. Hence, the operators $O_e$ and $O_m$ serve as `gauge' transformations; only the combinations of the two $O_eO_m$ transform a physical state, i.e. a $\ket{\phi^{+},\{x_i\}}$ state seen by the subsystem observer, to another. As such, the operator constraint of the physical states, now composed of only $O_eO_m$ operators, is the same as that in Section \ref{subSec:constraint}. So does TEE. The choice of the parameter $\phi$ here does not change the operator constraint and lead to no physical consequence, either. %Hence, TEE in closed EB cases is unambiguous and independent of all the parameters. 

\subsubsection{Open EB}

Now we are ready to deal with the case of $D(\mathbb{Z}_2)$ on a disk. Note that $\mathcal{H}^e=\mathcal{H}^{\{1,e\}|\{1,e\}}$ has $2^{n+1}$ norm-$1$ vectors and $\mathcal{H}^m=\mathcal{H}^{\{1,e\}|\{1,m\}}$ has $2^{n}$ norm-$\sqrt{2}$ vectors. The mapping between the two Hilbert spaces is two-to-one,
\begin{align}
\{\ket{\{x_i\};z_1,z_2}^e,\ket{\{x_i\};\tilde z_1,\tilde z_2}^e\}\leftrightarrow \ket{\{x_i\};\sigma,\sigma}^m;
\end{align}
With this map, the Hilbert space $\mathcal{H}^e\oplus\mathcal{H}^m$ can be decomposed into $2^{N-1}$ triplets. The general form of the reduced density matrix in this case is then
\be\label{Eq:generalopen}
\rho=\frac{1}{\Psi}\underbrace{\bordermatrix{
&\ket{\{x_i\};z_1,z_2}^e&\ket{\{x_i\};\tilde z_1,\tilde z_2}^e&\ket{\{x_i\};\sigma,\sigma}^m\cr
&p&c&d\cr
&c^*&p&f\cr
&d^*&f^*&2-2p}\oplus\cdots}_{2^{n}\textrm{ blocks}},
\ee
where $c,d,e$ are coherence terms to be added. To fulfil the advanced requirement of EM duality and vanish at least one eigenvalue for each block, it turns out that
\be
p=\frac{1}{2}; c=e^{i\phi}; d=\frac{\lambda e^{i\phi'}}{\sqrt{2}}; f=\frac{\lambda e^{i\phi-i\phi'}}{\sqrt{2}},
\ee
where $\lambda$, $\phi$, and $\phi'$ are some parameters that cannot be determined model-independently. The eigenvalues of each block is $\frac{1\pm\lambda}{2}$ and $0$. The eigenvectors for $\frac{1\pm\lambda}{2}$ are 
\be
\ket{\phi^\pm;\{x_i,z_1,z_2\}}\equiv\frac{1}{2}(\ket{\{x_i\};z_1,z_2}^e+e^{i\phi}\ket{\{x_i\},\tilde z_1,\tilde z_2}^e\pm 2^{\frac{1}{4}}e^{i\phi'}\ket{\{x_i\}, \sigma,\sigma}^m).
\ee
Verifying TEE gives $S_\textrm{TEE}=-\frac{1+\lambda}{2}\textrm{ln}(1+\lambda)-\frac{1-\lambda}{2}\textrm{ln}(1-\lambda)$. For TEE to be $-\frac{1}{2}$ln$2$, the parameter $\lambda$ is $\pm0.779944$. The parameters $p$ and $\lambda$ completely characterize the ambiguity in the open EB cases for $\mathbb{Z}_2$ topological order. These parameters suggest that there are extra degrees of freedom in factorizing the extended topological quantum field theory (TQFT) that underlies beneath the topological order with gapped boundaries, and the folding trick naturally chooses a specific values for $p$ and $\lambda$. While $p$ quantifies the probability of finding the states in a certain GBC, the physical meaning of $\lambda$ awaits to be clarified in future works.

%The $e^{i\phi}$ term adds coherence between the two states $\ket{\cdots,z_1,z_2}$ and $\ket{\cdots,\tilde z_1,\tilde z_2}$, which grants extra coherence between the two junctions. This term plays a crucial role in producing $S_{\textrm{TEE}}=\frac{1}{2}$ln$2$. In fact, if we set this term to zero, the eigenvalues are fixed to be $\lambda_+=3$ and $\lambda_-=1$, and the TEE is fixed to be $\frac{1}{4}$ln$\frac{1}{4}+\frac{3}{4}$ln$\frac{3}{4}$ no matter how we vary the other two coherence term. Therefore, to reproduce the compatibility with the folding trick, the correlation between two junctions should be stronger than that the constraint provides. We have already seen this stronger correlation when we fix the area-law term: the two junction points together behave as one EB point in the area-law term. 

We end our analysis by commenting on the operator constraint of the physical space. As in the case of closed EB, the choice of the two-to-one map is also not unique, and therefore some of the operators are `gauged', leaving one constraint on physical operators. According to Eq. \eqref{Eq:opencons}, the generating operator multiplies to $O_{z_1, z_2}$, whose explicit form will depend on $\lambda$ if restricted to the transformation between the physical states.

\section{Conclusion and outlook}

In this paper, we introduced a model-independent picture to understand the TEE. The TEE in general has three parts: a ln$d_x$ part due to the total charges of the EBs, where $x$ can be a bulk anyon or a boundary defect; a $-$ln$D$ part that characterized by the operator constraint; and a classical part due to the superposition of basis states. It turns out that in closed EB cases, the TEE is unambiguous and hence is good enough to characterize topological orders. The TEE of open EB cases, particularly the $-$ln$D$ part, depends on the EB condition and the choice of the mixed state. We introduced a general constructions of the reduced denstiy matrix for the $D(\mathbb{Z}_2)$ topological order in both closed EB cases and open EB cases. There are two key parameters in the construction: the probability $p$ of finding the EB in a certain GBC, and the coherence $\lambda$ between the two junction points. To reproduce the compatibility with the folding trick, on should adopt EM dual EB condition and set a certain value for $\lambda$. Though we have focused on the $D(\mathbb{Z}_2)$ topological order for simplicity, our analysis generalize easily to all Abelian topological orders. 

\begin{figure}[h!]
    \centering
    \includegraphics[scale=0.5]{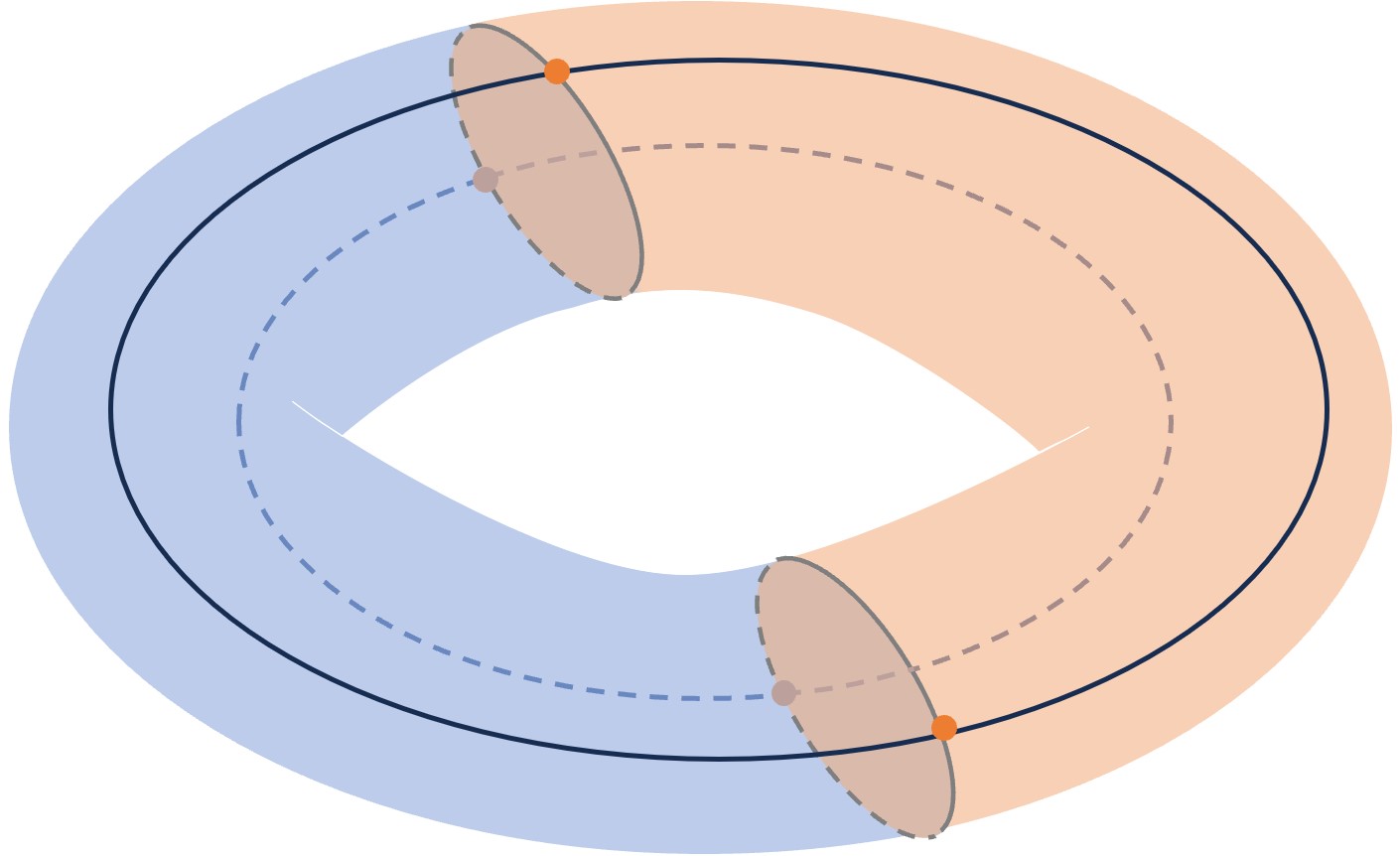}
    \caption{A composite system with topological orders (the blue region and the orange region) on a torus separated by a gapped domain wall. If an entanglement cut intersect with the domain wall, the intersection points will also exhibit non-trivial behavior.}
    \label{fig11:domainwall}
\end{figure}

In this paper, we focus on the topological orders that are describable by both lattice models and the CS theory. The primary reason for this restriction is because the mismatch is observed only in these topological orders. Nonetheless, our picture is model-independent, it is possible to generalize our analysis to non-Abelian topological orders with gapped boundaries. For non-Abelian topological orders, the operators are non-invertible and hence requires different treatment. Studying non-Abelian cases might require the notion of \textit{non-invertible symmetries}\cite{gaiotto_generalized_2015} and \textit{partial electric-magnetic duality}\cite{hu_electricmagnetic_2020}. Moreover, in the CS theory, the EBs are treated as gapless boundaries described by Ishibashi states. The discussion hints us that the gapped boundary and the gapless boundary of a topological order may be related in some way; it is likely that they differ by how the boundary points are coherent with each other. Moreover, the gapless CFT boundaries are \textit{gappable}\cite{levin_protected_2013}, studying the compatibility of gapped GBC and the folding trick will help us to further understand how CFT states are folded into gapped boundary states. Finally, more general cases, such as cases involve gapped domain walls\cite{zhao_characteristic_2023}, are also worth investigating. If the entanglement cut goes through the domain wall, as shown in Fig. \ref{fig11:domainwall}, the EB-domain wall junction will also have interesting properties.

\acknowledgments
YL thanks Ling-yan Hung, Yidun Wan, Hongyu Wang, Yanyan Chen, and Yu Zhao for helpful discussions. The author appreciates his supervisor Prof. Yidun Wan for supporting him via the General Program of Science and Technology of Shanghai No. 21ZR1406700, and Shanghai Municipal Science and Technology Major Project (Grant No.2019SHZDZX01). 
\appendix
\section{Review of properties of anyons}\label{appd:anyon}

We review the necessary fact about anyons that we will use in this paper. Interested readers are invited to read the review by Nayak\cite{nayak_nonabelian_2008} for a comprehensive introduction.

Anyons are fundamental excitations in topological orders. For a certain topological order, we label the anyon species as $\mathcal{C}=\{\alpha,\beta,\gamma,\cdots\}$. The anyons subject to fusion rules,
\be
\alpha\times \beta=\sum_\gamma N^\gamma_{\alpha\beta} \gamma,
\ee
where $N^\gamma_{\alpha\beta}$ are non-negative integers. Physically, fusing two anyons means if the two anyons $\alpha$ and $\beta$ are close enough to each other, they will behave like a superposition of $\gamma$ anyons. If we are in a state where a set of anyons fuse to $\gamma$, then we say that the \textit{total topological charge} of this set of anyons is $\gamma$. The trivial anyon $1$, also called the vacuum, is the identity of the fusion
\be
1\times \alpha=\alpha\times 1=\alpha.
\ee
The conjugate anyon $\bar\alpha$ of an anyon $\alpha$ is defined as $N^1_{\bar\alpha\alpha}\neq 0$. For an anyon $\alpha$, if for each $\gamma$ there is only one $\beta$ that satisfies $N^\gamma_{\alpha\beta}=1$ and all other $N^\gamma_{\alpha\beta'}=0$, the anyon $\alpha$ is called \textit{Abelian}. If all anyons in a topological order are Abelian, then the topological order is called an Abelian topological order. 

The Hilbert space of multiple anyons on a sphere is spanned by the possible fusion channels of these anyons. We can denote the basis states of the $n$-anyon Hilbert space using the following diagram,
\be\label{EqA:basis}
H_{n \textrm{ anyons}}=\textrm{span}\{\BLvert \basis \Brangle, \textrm{for all allowed } \{\beta_i\}\}
\ee
where $\beta_i\in \mathcal{C}$ are internal degrees of freedom, and we impose fusion rules on each vertex. One finds that the dimension of the Hilbert space for Abelian anyons is always $1$. 

The quantum dimension $d_\alpha$ of the anyon $\alpha$ can is the effective dimension of the single anyon Hilbert space in the thermodynamic limit. The quantum dimensions can be derived from the fusion rule,
\be\label{EqA:qdfusion}
d_\alpha d_\beta=\sum_{\gamma} N^\gamma_{\alpha\beta
} d_\gamma.
\ee
One immediately finds that $d_{\bar\alpha}=d_{\alpha}$ and $d_1=1$, and all Abelian anyons have quantum dimension $1$.

The most prominent property of the anyons is their non-trivial braiding statistics. Exchanging the position of two anyons will give one a unitary transformation on the internal Hilbert space of the anyons,
\be
\BLvert \braid \Brangle= \sum_{\beta'_i} B^{\alpha_i\alpha_{i+1}}_{\beta_i \beta'_i} \BLvert\unbraid\Brangle,
\ee
where $B$ stands for the transformation. For Abelian anyons, this unitary transformation is merely a phase factor. A crucial point is that we can detect anyons based on the non-trivial braiding statistics. Suppose we create two pairs of anyons $\alpha$, $\bar\alpha$, $\beta$, and $\bar\beta$ on a sphere. If we let the anyon $\beta$ wind around the anyon $\alpha$, and then annihilate the two pairs $\alpha$ and $\bar\alpha$, $\beta$ and $\bar\beta$, we will result into a matrix entry $S_{\alpha\beta}$. 
\be
S_{\alpha\beta}=\BLvert \smat\Brangle.
\ee
This matrix $S_{\alpha\beta}$ is an important observable that characterizes the topological order and is called the \textit{topological S-matrix}. Physically, one can use this AB phase to distinguish anyons: one can create an anyon pair, let them wind around the anyon to be recognized, and then annihilate the test anyons to acquire the amplitude of this process. Repeat this process for all $\alpha\in\mathcal{C}$ gives one a column in the S-matrix, which is enough to recognize the anyon.

\section{Norm of the boundary anyon states}\label{appd:bdrynorm}

\begin{figure}[h!]
    \centering
    \includegraphics[scale=0.45]{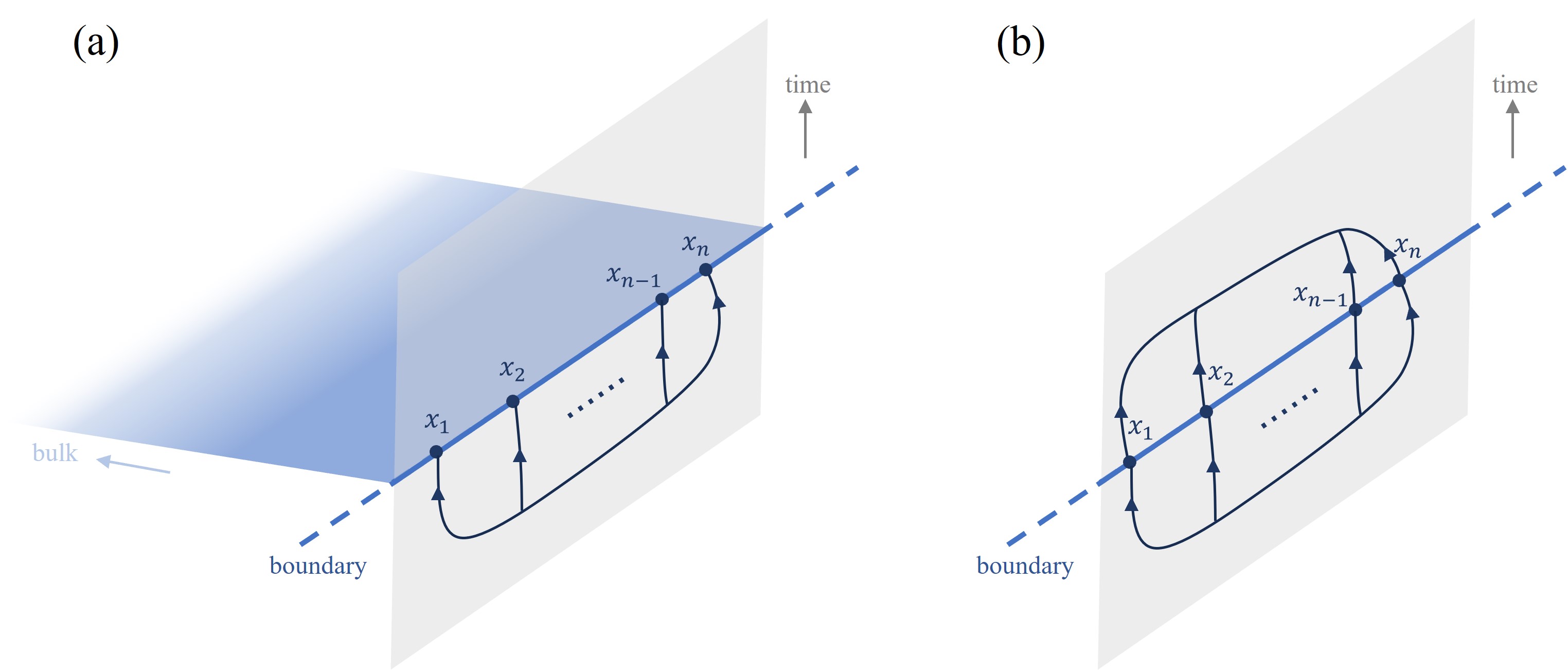}
    \caption{Caption}
    \label{figA1:bdrynorm}
\end{figure}

To understand the norm of the boundary anyon state, we consider the whole space-time picture as shown in Fig. (\ref{figA1:bdrynorm}).

Suppose we have $n$ anyons on the boundary. They were created at the time before the time slice we are focusing on. The worldline of these $n$ boundary anyons can be arranged as shown in the Fig. (\ref{figA1:bdrynorm}a). The basis of the $n$-anyon Hilbert space is also represented by these worldline diagrams, which is exactly the basis vectors we introduced in Eq. (\eqref{EqA:basis}). The Hermitian conjugate of a $n$-anyon state is conversely represented by the worldline of the $n$ anyons that annihilate with each other in the future, as shown in Fig. (\ref{figA1:bdrynorm}b). Taking the norm amounts to connecting the two diagrams and finding the amplitude of the diagram
\be
\bmm \loopamp \emm,
\ee
which is $\sqrt{d_{x_1}d_{x_2}\cdots d_{x_n}}$ for all states in this four-anyon Hilbert space.(See ref \cite{nayak_nonabelian_2008} for details).

\section{Verifying the normalization factor and the entanglement entropy}\label{appd:calculate}

We stick to the notation where $D$ is the total quantum dimension of the bulk topological order, and the lower latin alphabets $a,b,x$ represent for boundary anyons. We have $D=\sum_{a\in\mathcal{D}}d_a^2$, where $\mathcal{D}$ is the set of boundary anyons.

\begin{itemize}
\item \be
\sum_{\prod x_i=a} d_{x_1}d_{x_2}\cdots d_{x_n}=d_a D^{n-1}
\ee
\textit{Proof.}
We prove this equation by induction. For $n=1$ this equation holds trivially. Suppose the equation holds for a generic $n$; then, for the $n+1$ case,
\begin{align*}
&\sum_{\prod_{x_i}=a}d_{x_1}\cdots d_{x_n}d_{x_{n+1}}\\
=&\sum_{x_{n+1}}d_{x_{n+1}}\sum_{\prod_{x_i}^{i\leq n}=\bar x_{n+1} a}d_{x_1}\cdots d_{x_n},
\end{align*}
where the second summation in the second line runs over all possible combinations of $x_1\cdots x_n$ that fuses to any of the $c$'s in $\bar x_{n+1}\times a=N^{c}_{\bar x_{n+1} a} c$. Then,
\begin{align*}
&\sum_{\prod_{x_i}=a}d_{x_1}\cdots d_{x_n}d_{x_{n+1}}\\
=&\sum_{x_{n+1}}d_{x_{n+1}}N^{c}_{\bar x_{n+1} a}\sum_{\prod_{x_i}^{i\leq n}=c}d_{x_1}\cdots d_{x_n}\\
=&\sum_{x_{n+1}}d_{x_{n+1}}N^{c}_{\bar x_{n+1} a}d_c D^{n-1}\\
=&\sum_{x_{n+1}}d_{x_{n+1}} d_{x_{n+1}} d_{a} D^{n-1}\\
=&d_{a} D^{n},
\end{align*}
where in the third equality we used Eq. (\eqref{EqA:qdfusion}).

\item \be
\sum_{\prod x_i=a} \frac{d_{x_1}\cdots d_{x_n}}{d_a D^{n-1}} \textrm{ln}\frac{d_{x_1}\cdots d_{x_n}}{d_a D^{n-1}}=n\sum_{x\in\mathcal{D}}\frac{d_x^2}{D}\textrm{ln}\frac{d_x}{D} + \textrm{ln}\frac{D}{d_a}
\ee
\textit{Proof.}
\begin{align*}
&\sum_{\prod x_i=a} \frac{d_{x_1}\cdots d_{x_n}}{d_a D^{n-1}} \textrm{ln}\frac{d_{x_1}\cdots d_{x_n}}{d_a D^{n-1}}\\
=&\sum_{x_n}d_{x_n}^2 \left[\sum_{\prod^{n-1}_{i=1} x_i=\bar x_n a}\frac{d_{x_1}\cdots d_{x_{n-1}}}{d_ad_{x_n} D^{n-1}} \textrm{ln}d_{x_1}\cdots d_{x_n}\right]+\textrm{ln}\frac{1}{d_aD^{n-1}}\\
=&\sum_{x_n}d_{x_n}^2 \left[\sum_{\prod^{n-1}_{i=1} x_i=\bar x_n a}\frac{d_{x_1}\cdots d_{x_{n-1}}}{d_ad_{x_n} D^{n-1}} \textrm{ln}d_{x_1}\cdots d_{x_{n-1}} + \frac{1}{D}\textrm{ln}d_{x_n}\right]+\textrm{ln}\frac{1}{d_aD^{n-1}}\\
=&\sum_{x_n}d_{x_n}^2 \left[\sum_{\prod^{n-1}_{i=1} x_i=\bar x_n a}\frac{d_{x_1}\cdots d_{x_{n-1}}}{d_ad_{x_n} D^{n-1}} \textrm{ln}d_{x_1}\cdots d_{x_{n-1}}\right] + \sum_{x} \frac{d_x^2}{D}\textrm{ln}d_{x}+\textrm{ln}\frac{1}{d_aD^{n-1}}\\
=&\sum_{x_n}d_{x_n}^2 \left[\sum_{x_{n-1}}d_{x_{n-1}}^2\left(\sum_{\prod^{n-2}_{i=1} x_i=\bar x_n\bar x_{n-1} a}\frac{d_{x_1}\cdots d_{x_{n-2}}}{d_ad_{x_n}d_{x_{n-2}} D^{n-1}} \textrm{ln}d_{x_1}\cdots d_{x_{n-2}}\right)+\frac{1}{D^2}\textrm{ln}d_{x_{n-1}}\right] \\
&\hspace{30pt}+ \sum_{x} \frac{d_x^2}{D}\textrm{ln}d_{x}+\textrm{ln}\frac{1}{d_aD^{n-1}}\\
=&\sum_{x_n,x_{n-1}}d_{x_n}^2d_{x_{n-1}}^2\left[\sum_{\prod^{n-2}_{i=1} x_i=\bar x_n\bar x_{n-1} a}\frac{d_{x_1}\cdots d_{x_{n-2}}}{d_ad_{x_n}d_{x_{n-1}} D^{n-1}} \textrm{ln}d_{x_1}\cdots d_{x_{n-2}}\right]+2\sum_{x} \frac{d_x^2}{D}\textrm{ln}d_{x}+\textrm{ln}\frac{1}{d_aD^{n-1}}\\
&\vdots\\
&\hspace{5pt}\textrm{repeat }n\textrm{ times}\\
&\vdots\\
=&n\sum_{x} \frac{d_x^2}{D}\textrm{ln}d_{x}+\textrm{ln}\frac{1}{d_aD^{n-1}}=n\sum_{x} \frac{d_x^2}{D}\textrm{ln}\frac{d_{x}}{D}+\textrm{ln}\frac{D}{d_a}.
\end{align*}
\end{itemize}

\section{The local term due to inhomogeneous EB}\label{appd:localterm}
We focus on the $D(\mathbb{Z}_2)$ toric code on a sphere. The Eq. (\eqref{Eq:statepartition}) is not the unique choice of how the global state is constructed by the states in the subsystem Hilbert spaces. 
The ground state of the topological order on a sphere can be instead decomposed as, for example,
\begin{align*}
\ket{\psi}=\frac{1}{\sqrt{\Psi}}&\sum_{\prod x_i=1} \sqrt{\lambda}(\ket{x_{pq}}+e^{i\phi/2}\ket{\tilde x_{pq}})_{\textrm{A}}(\ket{\tilde x_{pq}}+e^{i\phi/2}\ket{x_{pq}})_{\textrm{B}}\\
&+\sqrt{(1-\lambda)}(\ket{x_{pq}}-e^{i\phi/2}\ket{\tilde x_{pq}})_{\textrm{A}}(\ket{\tilde x_{pq}}-e^{i\phi/2}\ket{x_{pq}})_{\textrm{B}},
\end{align*}
where $\ket{x_{pq}}\equiv \ket{x_1,\cdots,x_p,\cdots,x_q,\cdots,x_n}$ and $\ket{\tilde x_{pq}}\equiv \ket{x_1,\cdots,\tilde x_p,\cdots,\tilde x_q,\cdots,x_n}$ and $\lambda$ is a real number $\lambda\in[0,1]$.
This decomposition assumes extra coherence between the two points $x_p$ and $x_q$. The reduced density matrix then becomes
\be
\rho=\frac{1}{\Psi}\underbrace{\bordermatrix{
&\ket{\{x_{pq}\}}&\ket{\{\tilde x_{pq}\}}\cr
&1&\lambda e^{i\phi}\cr
&\lambda e^{-i\phi}&1}\oplus\cdots}_{2^{n-1}\textrm{ blocks}}.
\ee
Verifying the TEE of this reduced density matrix gives 
\be
S=(n-1) \textrm{ln} 2 -\frac{1+\lambda}{2}\textrm{ln}\frac{1+\lambda}{2} -\frac{1-\lambda}{2} \textrm{ln}\frac{1-\lambda}{2}.
\ee
There is an extra sub-leading term in the entanglement entropy, as we mentioned in the main text.

%%%%%%%%%%%%%%%%%%%
%%%%%%%%%%%%%%%%%%%
%%%%%%%%%%%%%%%%%%%

\bibliographystyle{apsrev4-2}
\bibliography{Topological}
\end{document}